\documentclass[12pt]{article}
\pdfoutput=1
\usepackage{putex}
\usepackage{amsmath}
\usepackage{amssymb}
\usepackage{graphicx}
\usepackage{epstopdf}
\usepackage{enumerate}
\usepackage{cite}
\usepackage{tensor}
\usepackage{slashed}
\usepackage{feynmf}
\usepackage{hyperref}
\usepackage{bbold}
\usepackage{dsfont}
\usepackage{mathrsfs}
\usepackage{caption}
\usepackage{subcaption}

\usepackage{youngtab}
\usepackage{amsfonts}
\usepackage{braket}

\newcommand{\lp}{L_1}
\newcommand{\lm}{L_{-1}}

\newcommand{\cz}{c_0}
\newcommand{\cp}{c_1}
\newcommand{\cm}{c_{-1}}

\newcommand{\SL}[1]{SL(#1)}
\newcommand{\tC}{\tilde{C}}
\newcommand{\Rc}{\mathcal{R}}

\newcommand{\al}[2]{\begin{align}\label{#1}\begin{split}#2\end{split}\end{align}}

\newcommand{\qk}{q^{\scriptscriptstyle (k)}}
\newcommand{\qbk}{\bar{q}_{\scriptscriptstyle (k)}}
\newcommand{\qi}{q^{\scriptscriptstyle (i)}}
\newcommand{\qbi}{\bar{q}_{\scriptscriptstyle (i)}}
\newcommand{\qo}{q^{\scriptscriptstyle (1)}}
\newcommand{\qbo}{\bar{q}_{\scriptscriptstyle (1)}}
\newcommand{\qt}{q^{\scriptscriptstyle (2)}}
\newcommand{\qbt}{\bar{q}_{\scriptscriptstyle (2)}}
\newcommand{\tq}{\tilde{q}}
\newcommand{\qth}{q^{\scriptscriptstyle (3)}}
\newcommand{\qbth}{\bar{q}_{\scriptscriptstyle (3)}}
\newcommand{\qf}{q^{\scriptscriptstyle (4)}}
\newcommand{\qbf}{\bar{q}_{\scriptscriptstyle (4)}}

\newcommand{\F}{{\cal F}}
\newcommand{\hb}{\tilde{h}}

\newcommand{\lop}{\lambda^{\prime}_1}
\newcommand{\lo}{\lambda_1}
\newcommand{\ltp}{\lambda^{\prime}_2}
\newcommand{\lt}{\lambda_2}
\newcommand{\mup}{\mu'}

\newcommand{\Smalldots}{\text{{\small  $\dots$}}}

\usepackage[utf8x]{inputenc}
\usepackage{braket}
\usepackage{titlesec}
\usepackage{fixltx2e}
\usepackage{cleveref}
\hypersetup{breaklinks}

\usepackage[usenames,dvipsnames,svgnames,table]{xcolor}

\def\l{\lambda}

\def\Tt{\tilde{T}}

\def\p{\partial}

\newcommand {\be} {\begin {equation}}
\newcommand {\ee} {\end {equation}}
\newcommand{\bea}{\begin{eqnarray}}
\newcommand{\eea}{\end{eqnarray}}

\def\zb{\overline{z}}
\def\wb{\overline{w}}
\def\rt{\rightarrow}

 \newmuskip\pFqmuskip

\newcommand*\pFq[6][8]{%
  \begingroup 
  \pFqmuskip=#1mu\relax
  \mathcode`\,=\string"8000
  \begingroup\lccode`\~=`\,
  \lowercase{\endgroup\let~}\pFqcomma
  {}_{#2}F_{#3}{\left[\genfrac..{0pt}{}{#4}{#5};#6\right]}%
  \endgroup
}
\newcommand{\pFqcomma}{\mskip\pFqmuskip}

\makeatletter
\renewcommand{\@maketitle}{
\newpage
 \begin{center}%
  {\large\bfseries \@title \par}%
 \end{center}%
 \par} \makeatother

\numberwithin{equation}{section}

\titleformat*{\section}{\large\bfseries}

\begin{document}

\institution{UCLA}{Department of Physics and Astronomy, University of California, Los Angeles, CA 90095, USA}

\title{Holographic conformal blocks  \\from interacting Wilson lines}

\authors{Mert Besken\worksat{\UCLA}, Ashwin Hegde\worksat{\UCLA}, Eliot Hijano\worksat{\UCLA},  Per Kraus\worksat{\UCLA}}

\abstract{We present a simple prescription for computing conformal blocks and correlation functions holographically in AdS$_3$  in terms of Wilson lines merging at a bulk vertex. This is shown to reproduce global conformal blocks and heavy-light Virasoro blocks.  In the case of higher spin theories  the space of vertices is in one-to-one correspondence with the space of ${\cal W}_N$ conformal blocks, and we show how the latter are obtained by explicit computations.    }

\date{}

\maketitle
\setcounter{tocdepth}{2}
\tableofcontents

\section{Introduction}

This paper continues a program aimed at determining the AdS gravity description of conformal blocks. For previous work see \cite{Fitzpatrick:2014vua,Hijano:2015rla,deBoer:2014sna,Hijano:2015zsa,
Hijano:2015qja,Hegde:2015dqh,Fitzpatrick:2015zha,Alkalaev:2015wia,
Banerjee:2016qca,Bhatta:2016hpz}. The conformal block decomposition of correlation functions, combined with the constraints of unitarity and crossing symmetry,  is a powerful nonperturbative framework in which to study strongly interacting conformal field theories \cite{Rychkov:2016iqz,showk,Qualls:2015qjb,Simmons-Duffin:2016gjk}. It has also proven to be very effective in elucidating the AdS/CFT correspondence, in particular the emergence of local physics in the bulk \cite{Heemskerk:2009pn,Heemskerk:2010ty,Penedones:2010ue,ElShowk:2011ag,Fitzpatrick:2012cg,
Fitzpatrick:2012yx,Fitzpatrick:2014vua,Jackson:2014nla}.

To push this program forward it is very useful to have in hand bulk AdS representations of conformal blocks.  In \cite{Hijano:2015zsa} it was shown that global conformal blocks with external scalar operators have a simple bulk representation in terms of ``geodesic Witten diagrams". This refers to a tree level exchange Witten diagram with a pair of cubic vertices, except that the vertices are not integrated over all of AdS, but  only over geodesics connecting the boundary points hosting the external operators.  This result leads to a strikingly simple procedure for expanding the full Witten diagram in conformal blocks.

In the case of AdS$_3$/CFT$_2$ the story is especially rich since the global conformal algebra is enhanced to an infinite dimensional algbebra, namely Virasoro or something larger, such as a ${\cal W}$-algebra.  Here one focusses on the regime of large central charge, since this is the regime where the bulk becomes classical.  In \cite{Fitzpatrick:2014vua,Hijano:2015rla,Hijano:2015qja,
Alkalaev:2015wia,Banerjee:2016qca} it was shown that heavy-light Virasoro blocks (defined by scaling some operator dimensions with $c$, while keeping others fixed)  are reproduced by geodesic Witten diagram operators, now not in pure AdS$_3$ but in a new geometry produced by backreaction from the heavy operators.

Conformal blocks for ${\cal W}$-algebras are relevant to the recent interest in higher spin AdS/CFT dualities.   In particular, Gaberdiel and Gopakumar \cite{Gaberdiel:2010pz} proposed to consider the minimal model cosets
\be
{{{\rm SU}(N)_k\oplus {\rm SU}(N)_1}\over{{\rm SU}(N)_{k+1}}}
\ee
in the 't Hooft limit $k, N \rt \infty$ with $\lambda = N/(N+k)$ fixed. This was argued to be holographically dual to the higher spin theory of Prokushkin and Vasiliev \cite{Prokushkin:1998bq}.  The theory in the 't Hooft limit has left and right moving ${\cal W}_{\infty}(\lambda)$ algebras \cite{Gaberdiel:2011wb,Gaberdiel:2012ku}.  These are nonlinear algebras with an infinite tower of conserved currents.  It is then of interest to know the corresponding conformal blocks, but these are rather challenging to obtain directly  on account of the complexity of the algebra.

At fixed $N$ the algebras are ${\cal W}_N$, with conserved currents of spins $s=2, \ldots N$.  One of the main results of this paper is to provide a very simple bulk prescription for the conformal blocks of these algebras in the large $c$ limit.  Furthermore, this can be used as a backdoor approach for obtaining (some of) the ${\cal W}_{\infty}(\lambda)$ blocks, as this can be achieved by the analytic continuation $N\rt -\lambda$; see \cite{Campoleoni:2013lma,Hegde:2015dqh} for examples of this approach.  We also note that upon setting $N=2$ the conformal blocks are those of the Virasoro algebra.

The setup we use can be motivated as follows.  We note that the central charge of the coset theory is
\be
c=(N-1)\left(    1-{{N(N+1)}\over {(N+k)(N+k+1)}} \right)
\ee
To take $c \rt \infty$ at fixed $N$ we can take the limit $k \rt -N-1$, dubbed the ``semiclassical limit" in \cite{Perlmutter:2012ds}.  The negative value of $k$ results in a non-unitary theory, manifested for example by negative dimension primaries in the spectrum.  As a result, this limit does not provide a healthy example of the AdS/CFT correspondence in Lorentzian signature (see also \cite{Perlmutter:2016pkf} for related discussion).  However, as noted above it does act as a useful stepping stone for obtaining results in the unitary 't Hooft limit via analytic continuation in $N$.  It is also of interest --- perhaps as a warmup example --- as a very explicit and tractable setup where many details of AdS/CFT an be worked out.

For example, all coset primaries in this limit can be identified in the bulk, at least below the black hole threshold.   The bulk description is in terms of SL(N)$\times \widetilde{{\rm SL(N)}}$ Chern-Simons theory coupled to matter.  Coset primaries are labelled by a pair of SL(N) highest weights, $(\Lambda_+,\Lambda_-)$.  These are highest weights of finite dimensional representations of SL(N).  Primaries of the form $(0,\Lambda_-)$ have scaling dimension $\Delta\sim c$; they are ``heavy" operators, and are described in the bulk by flat SL(N)$\times \widetilde{{\rm SL(N)}}$ connections \cite{Castro:2011iw}.   On the other hand primaries of the form $(\Lambda_+,0)$ have $\Delta \sim O(1)$; these light operators are described by perturbative matter in the bulk.  The general $(\Lambda_+,\Lambda_-)$ is then described by light matter fields propagating in  the heavy classical background \cite{Perlmutter:2012ds,Hijano:2013fja}.

The main result of this work is a simple and usable expression for computing correlators of these operators, significantly extending previous work.  Let us first consider the case of $n$ light operators.  The correlator is described by $n$ bulk-to-boundary propagators meeting at an n-point vertex, according to the following rules.  Each light operator corresponds to a representation of SL(N)$\times \widetilde{\rm SL(N)}$ with highest weight state $|{\rm hw}\rangle_i|{\rm \widetilde{hw}}\rangle_i$, $i=1, \ldots n$.  We then attach a Wilson line to each such state\footnote{ Wilson lines first made an appearance in these theories in the context of entanglement entropy \cite{deBoer:2013vca,Ammon:2013hba}, and have appeared more recently as a probe of black hole solutions \cite{Castro:2016ehj}. }, emanating from the associated boundary point $x_i$ to a point in the bulk, $Pe^{\int_{x_i}^{x_b} A } Pe^{\int_{x_i}^{x_b} \tilde{A}}$. Since the connections are flat, the choice of path does not matter. The bulk vertex located at $x_b$ is defined by choosing a singlet state $|S\rangle$  in the tensor product of representations corresponding to the boundary operators.  In general, there are many choices for such singlet states, and as we discuss below these are in one-to-one correspondence with conformal blocks, as can be seen by taking the tensor product of pairs of operators, and then combining terms in the product into singlets.
With these ingredients in hand, the correlator is\footnote{An equivalent formula was proposed and studied in the $N=2$ context in the recent paper \cite{Bhatta:2016hpz}, which appeared while this work was in progress.}
\be\label{master}
G_S(z_i,\zb_i) = \langle S|\prod_{i=1}^n Pe^{\int_{x_i}^{x_b} A } |{\rm hw}\rangle_i  Pe^{\int_{x_i}^{x_b} \tilde{A}} |{\rm \widetilde{hw}}\rangle_i
\ee
The correlator is independent of the choice of $x_b$, as seen by noting that changing $x_b$ just introduces a group element that acts on the singlet state as the identity.
To include heavy primaries $(0,\Lambda_-)$ we still use (\ref{master}) but now with $(A,\tilde{A})$ taken to be the flat connection representing the heavy background; this is especially simple in the case of two heavy operators in conjugate representations, which is all that we consider in this paper, while more generally one needs to solve a nontrivial monodromy problem \cite{deBoer:2014sna}.   The general $(\Lambda_+,\Lambda_-)$ primary is included by taking the location of a light $(\Lambda_+,0)$ primary to coincide with the insertion point of the heavy $(0,\Lambda_-)$ primary.

Our master formula (\ref{master}) reduces the problem of computing correlators to computing SL(N) matrix elements. We will verify that we correctly reproduce various known results for four-point functions. First, it's easy to see that we reproduce all previous results \cite{Fitzpatrick:2014vua,Hijano:2013fja,deBoer:2014sna,Hegde:2015dqh} for vacuum blocks.  Setting $N=2$ and taking all operators to be light we obtain the well known formula for global conformal blocks.  Taking
two operators to be heavy we correctly reproduce heavy-light Virasoro blocks.   For $N=3$ with four light operators we obtain the result for ${\cal W}_3$ blocks found in \cite{Fateev:2011qa}.  Allowing $N$ to be arbitrary and taking light operators in the fundamental and anti-fundamental representations we reproduce previous results derived using the Coulomb gas formalism \cite{Papadodimas:2011pf}.  In all these cases, the primaries we consider have negative scaling dimension, due to the underlying non-unitarity.  However, it is easy to analytically continue to positive dimensions and obtain results in the unitary regime.

In our construction, each choice of singlet state yields a correlator.  As we already mentioned, there is a natural basis for such singlet states that gives a one-to-one correspondence with conformal blocks.  The general correlator is then a general sum over products of left and right moving conformal blocks.  Of course, any particular theory will lead to particular coefficients in this sum. For example, this would be the case if we had derived (\ref{master}) starting from, say, a Lagrangian.  In principle, it should be possible to start from the equations of Prokushkin and Vasiliev and derive the precise correlators that reproduce those of the coset theory, and it would be very interesting to do so.

Apart from a relation to any particular CFT, what the Wilson line approach does is allow one to compute conformal blocks for operators in degenerate representations of the chiral algebra. For example, in the $N=2$ Virasoro case the dimensions of degenerate primaries are given by the famous Kac formula, $h=h_{r,s}(c)$.  As $c\rightarrow \infty$,
\be
h_{1,s}(c)  =  -{s-1\over 2}+ O(1/c)~,\quad h_{r,1}(c) = -{r^2-1\over 24}c+  O(c^0)~.
\ee
Light operators of dimension $h_{1,s}$ will be seen to be described by Wilson lines in the spin $j=(s-1)/2$ representation of SL(2), while heavy operators of dimension $h_{r,1}$ correspond to flat connections whose holonomy around the boundary has winding number $r$. Since minimal models are built up out of degenerate representations, we can use Wilson lines and flat connections to compute correlators in these theories.

\section{Correlation functions: general formulation}\label{sec:GeneralFormulation}

In this section we motivate and present our general expression for correlators and conformal blocks, and illustrate with a few simple examples.

\subsection{Preliminaries}

We will be dealing with the group SL(N)$\times \widetilde{\rm SL(N)}$.  The generators of the principally embedded SL(2) are denoted as\footnote{These are typically denoted as $L_i$, but we reserve $L_i$ for SL(2) matrices in the N dimensional defining representation of SL(N).}  $T_i$, $i=-1, 0, 1$ and obey $[T_i,T_j]=(i-j)T_{i+j}$. We similarly introduce $\Tt_i$ generators for $\widetilde{\rm {SL(2)}}$.

Each primary ${\cal O}_i$ will be associated with a finite dimensional representation $({\cal R}_i,\tilde{\cal R}_i)$ of SL(N)$\times$$\widetilde{{\rm SL(N)}}$. We denote the highest weight state in this representation as $|{\rm hw}\rangle_i | \widetilde{\rm hw}\rangle_i$, where the notion of highest weight is determined by maximizing the eigenvalues of $T_0$ and $\tilde{T}_0$. The scaling dimensions of these operators $(h_i,\tilde{h}_i)$ are determined by the highest weights:
\be
T_0|{\rm hw}\rangle_i=-h_i|{\rm hw}\rangle_i~,\quad  \Tt_0| \widetilde{\rm hw}\rangle_i=-\tilde{h}_i| \widetilde{\rm hw}\rangle_i~.
\ee

The connections for SL(N) and $\widetilde{{\rm SL(N)}}$ are denoted $A$ and $\tilde{A}$ respectively.  AdS$_3$ with planar boundary is described by
\be
A= e^\rho T_1 dz +T_0 d\rho~,\quad \tilde{A} = e^\rho \Tt_1 d\zb - \Tt_0 d\rho
\ee
As is standard, a gauge transformation can be performed to effectively remove all reference to the radial coordinate $\rho$, so that we work with $a = T_1 dz$ and $\tilde{a}= \Tt_1 d\zb$.  More general backgrounds are obtained by replacing the generators $T_1$ and $\Tt_1$ by other group generators, and we describe these later as needed.  More details can be found in any number of references; e.g. \cite{Campoleoni:2010zq,Ammon:2012wc}

\subsection{Correlators}

We start out by considering the correlation function of n primary operators on the plane
\be
G(x_i)=\langle {\cal O}_1(x_1)\ldots {\cal O}_n(x_n)\rangle~.
\ee

An n-point correlator is built out of n bulk-to-boundary propagators meeting at a bulk vertex located at the point $(\rho_b, z_b, \zb_b)$. Since results will not depend on the choice of $\rho_b$ we suppress it throughout.  Neither will results depend on the choice of $(z_b,\zb_b)$, but intermediate computations simplify for certain choices, so dependence on these quantities will be retained.  The bulk-to-boundary propagator emanating from boundary point $(z_i,\zb_i)$ is
\be\label{aa}
 P e^{\int_{x_i}^{x_b}a}|{\rm hw}\rangle_i e^{\int_{x_i}^{x_b}\tilde{a}} | \widetilde{\rm hw}\rangle_i=e^{z_{bi}T^{(i)}_1} |{\rm hw}\rangle_i e^{\zb_{bi}\Tt^{(i)}_1} | \widetilde{\rm hw}\rangle_i~.
\ee
where $z_{bi}=z_b-z_i$ and $\zb_i=\zb_b-\zb_i$.  Note that (\ref{aa}) is a state in the representation $({\cal R}_i,\tilde{\cal R}_i)$.

The bulk vertex is defined by choosing a singlet state in the tensor product $({\cal R}_1,\tilde{\cal R}_1)\otimes \ldots \otimes ({\cal R}_n,\tilde{\cal R}_n)$.  As discussed below, a particular basis for such singlet states corresponds to a basis of conformal blocks in which to expand the correlation function. Certain linear combinations of these basis states can then be used to construct a correlation function obeying crossing symmetry.  Given a choice of singlet state $|S\rangle$, the corresponding correlator is given by the matrix element

\be\label{ab}
G_S(z_i,\zb_i) = \langle S| \prod_{i=1}^n  e^{z_{bi}T^{(i)}_1} |{\rm hw}\rangle_i e^{\zb_{bi}\Tt^{(i)}_1} | \widetilde{\rm hw}\rangle_i~.
\ee
We show below that this object transforms correctly under the global conformal group.

It is natural to adopt a basis of singlet states which factorize as $|S\rangle = |s\rangle|\tilde{s}\rangle$.  The general correlation function is then a sum of holomorphically factorized terms,
\be
\label{ac}
G(x_i) = \sum_{s\tilde{s}} A_{s\tilde{s}}w_s(z_i)\tilde{w}_{\tilde{s}}(\zb_i)~,
\ee
with
\be
\label{ad}
w_s(z_i) = \langle s| \prod_{i=1}^n  e^{z_{bi}T^{(i)}_1} |{\rm hw}\rangle_i~,\quad\text{and}\quad \tilde{w}_{\tilde{s}}(\zb_i) = \langle \tilde{s}| \prod_{i=1}^n  e^{\zb_{bi}\Tt^{(i)}_1} |\widetilde{{\rm hw}}\rangle_i.
\ee
Once we have computed $w_s(z_i)$ the corresponding result for $\tilde{w}_{\tilde{s}}(\zb_i)$ follows by making obvious replacements.

We now note a few key properties satisfied by $w_s(z_i)$.   First, we establish that the expresssion in (\ref{ad}) is independent of the choice of bulk point $z_b$.  Suppose that instead of $z_b$ we place the vertex at $z_{b'}$; this gives back the same result:
\be w'_s(z_i) = \langle s| \prod_{i=1}^n  e^{z_{b'i}T^{(i)}_1} |{\rm hw}\rangle_i=\langle s|\prod_{i=1}^n e^{z_{b'b}T^{(i)}_1} \prod_{i=1}^n  e^{z_{bi}T^{(i)}_1} |{\rm hw}\rangle_i=w_s(z_i) ~,
\ee
where we used the fact that $\langle s|$ is a singlet, and hence invariant under the action of the group element $\prod_{i=1}^n e^{z_{b'b}T^{(i)}_1}$.

A similar argument explains why we do not have to consider any additional ``exchange" type diagrams in addition to the ``contact" diagram defined above.  An exchange diagram would have bulk vertices connected by bulk-to-bulk propagators.  But since the location of bulk vertices is arbitrary, we can always choose to move them all to a single point, in which case the bulk-to-bulk propagators are absent, and we simply recover a contact diagram.   The completeness of contact diagrams will be corroborated by the fact that these will be seen to yield a complete set of conformal blocks, out of which any correlator can be assembled.

We next establish that $w_s(z_i)$ transforms as it should under conformal transformations, namely
\be\label{ae}
w_s(z'_i) = \left[\prod_{i=1}^n \left({\p z'_i \over \p z_i}\right)^{-h_i} \right] w_s(z_i)~,\quad z'_i = {az_i+b \over cz_i+d}~.
\ee
We do this by applying a gauge transformation that acts as $z_i \rt z'_i$. The details are given in appendix \ref{appconf}.

While our main focus will be on 4-point functions, let us first illustrate by considering the computation of 2-point and 3-point functions.   Given (\ref{ae}), the dependence on $z$ is guaranteed to come out correctly in these cases, but verifying this is a useful warmup.

For the 2-point function, in order to construct a singlet state we need that the representations ${\cal R}_1$ and ${\cal R}_2$ be conjugates of each other.  In particular, this implies the familiar fact that the 2-point function vanishes unless the two operators have the same scaling dimension.   We use the freedom to choose $z_b$ arbitrarily to set $z_b=z_2$, which yields
\be\label{af}
w_s(z_1,z_2) = \langle s|e^{-z_{12}T^{(1)}_1} |{\rm hw}\rangle_1 |{\rm hw}\rangle_2~.
\ee
The singlet state is $|s\rangle = |-{\rm hw}\rangle_1 |{\rm hw}\rangle_2 + \ldots$.  The omitted terms contain states other than $|{\rm hw}\rangle_2$, but it's clear from (\ref{af}) that these won't contribute, and so
\be\label{ag}
w_s(z_1,z_2) = \langle -{\rm hw} | e^{-z_{12}T^{(1)}_1}| {\rm hw}\rangle_1 = {C\over (z_{12})^{2h}}~,
\ee
for some constant $C$.  To arrive at (\ref{ag}) we just used that the highest weight has $T_0$ eigenvalue $-h$, together with the fact that $T_1$ lowers the weight by one unit, to note that the only contribution comes from picking out the $-2h$ power from the expansion of the exponential.   The result (\ref{ag}) is of course the one dictated by conformal invariance.

We now turn to the three point function.  For this to be nonzero we need that ${\cal R}_1 \otimes {\cal R}_2 \otimes {\cal R}_3$ contains a singlet.  Although ${\cal R}_{i}$ are representations of SL(N) with highest weights $-h_i$, for the purposes of this computation we can take them to be representations of SL(2) of spin $j_i =-h_i$, and the singlet to be the SL(2) singlet built out of these three representations.  The reason is that in (\ref{ad}) we are acting with SL(2) group elements on the highest weight states, and these can only yield states in the same SL(2) representation.  That is, terms in the SL(N) singlet containing SL(2) spins different from $j_i$ yield no contribution.  With this in mind, the singlet is given by the Wigner 3j symbol as
\be
|s\rangle = \sum_{m_1,m_2,m_3} \left(\begin{array}{ccc}j_1 & j_2& j_3 \\ m_1 & m_2 &m_3 \end{array} \right)|j_1m_1\rangle|j_2m_2\rangle|j_3m_3\rangle~.
\ee
Using our freedom to choose the location of the bulk vertex, we take $z_b=z_1$, and note that this implies that only the term $m_1=j_1$ in the sum contributes. The three point function is
\bea
&&w_s(z_1,z_2,z_3)
\cr
&&= \sum_{m_1,m_2,m_3} \left(\begin{array}{ccc}j_1 & j_2& j_3 \\ m_1 & m_2 &m_3 \end{array} \right)\langle j_1m_1|e^{z_{b1}T_1^{(1)}}|j_1j_1\rangle \langle j_2m_2|e^{z_{b2}T_1^{(2)}}|j_2j_2\rangle\langle j_3m_3|e^{z_{b3}T_1^{(3)}}|j_3j_3\rangle~.  \cr &&
\eea
The sum can be evaluated using the known expression for the Wigner 3j symbol.  Alternatively, we can work in terms of tensors.  The latter approach generalizes more readily to our four-point computations, and in appendix \ref{appthreepoint} we show that this yields
\be\label{threepoint}
w_s(z_1,z_2,z_3) = { C(j_1,j_2,j_3)\over z_{12}^{h_1+h_2-h_3} z_{13}^{h_1+h_3-h_2} z_{23}^{h_2+h_3-h_1} }~,
\ee
where $C(j_1,j_2,j_3)$ is nonzero provided the product of the three representations contains a singlet.  Again, this is the standard result dictated by conformal invariance.

\section{Four-point functions}\label{sec:4pt}
This paper focuses mainly on the study of four-point functions of primary operators on the plane.
\begin{equation}
G(x_i)=\langle {\cal O}_1(x_1) {\cal O}_2(x_2) {\cal O}_3(x_3) {\cal O}_4(x_4)\rangle ~.
\end{equation}
As in the previous section, each primary corresponds to the highest weight state of an irreducible representation of SL(N)$\times$$\widetilde{{\rm SL(N)}}$ that we denote $({\cal R}_i,\tilde{\cal R}_i)$. In the following subsections we review the conformal block decomposition of four-point functions, we explain the construction of conformal blocks through the assembly of singlets, and we discuss restrictions due to crossing symmetry.

\subsection{Conformal block decomposition}

We now quickly review the conformal block decomposition of four-point correlators on the plane.  The correlator is expressed as a sum of conformal partial waves (CPWs), each of which corresponds to inserting a projector onto a single representation of the relevant symmetry algebra,
\be
\langle {\cal O}_1(x_1){\cal O}_2(x)P_P {\cal O}_3(x_3){\cal O}_4(x_4)=C^P_{12}C^P_{34}W_P(x_i) ~.
\ee
The projection operator $P_P$ projects onto the space of states in a representation labelled by the primary operator ${\cal O}_P$.  Pulling out the OPE coefficients renders $W_P(x_i)$ an object that is completely determined by symmetry, and in terms of which the full correlator is expanded as
\be
G(x_i) =\sum_P C^P_{12}C^P_{34} W_P(x_i)~.
\ee
Since the symmetry algebra factorizes into commuting left and right moving algebras, the same is true of the CPWs,
\be
W_P(x_i) = w_p(z_i) \tilde{w}_{\tilde{p}}(\zb_i)~.
\ee
Invariance under the global conformal group allows us to reduce the dependence to
\be \label{block}
 w_p(z_i)=\left({z_{24}\over z_{14}}\right)^{h_{12}}\left({z_{14}\over z_{13}}\right)^{h_{34}}\left({z_{34}\over z_{13}}\right)^{h_1+h_2}{g_p(z)\over z_{24}^{h_1+h_2}z_{34}^{h_3+h_4}}~,
\ee
where $h_{ij} \equiv h_i-h_j$,  $z_{ij} \equiv z_i-z_j$, and $z$ is the conformally invariant cross ratio
\be
z={z_{12}z_{34}\over z_{13}z_{24}}~.
\ee
The analogous result holds for $\tilde{w}_{\tilde{p}}(\zb_i)$ upon making the obvious substitutions.  We note that $g_p(z)$ depends on the quantum numbers of the primary operators appearing in the correlation function as well as those of the exchanged primary.

Another way to express the above is to use conformal invariance to set $x_1=z$, $x_2=0$, $x_3=\infty$ and $x_4=1$.   We then have \footnote{The prefactor in \eqref{block} was chosen such that the $w_p$ reduce to $g_p$ for pairwise identical operators at these distinguished positions. In the sections to follow we will assume that the prefactor has been chosen so.}
\begin{align}\label{block2}
\begin{split}
&\langle {\cal O}_1(z,\zb){\cal O}_2(0,0)P_P{\cal O}_3(\infty,\infty) {\cal O}_4(1,1)\rangle \cr
&\quad\quad\quad\quad =C^P_{12}C^P_{34}\left[(1-z)^{\vphantom{\tilde{h}}h_{34}-h_{12}}g_p(z)\right]\left[(1-\zb)^{\tilde{h}_{34}-\tilde{h}_{12}}\tilde{g}_{\tilde{p}}(\zb)\right]
\end{split}
\end{align}
where ${\cal O}_3(\infty,\infty)= \lim_{x_3\rt \infty}z^{2h_3}\zb^{2\tilde{h}_3}{\cal O}_3(x_3)$ inside the correlator. The form of $g_p(z)$ depends on what symmetry algebra is controlling the conformal block decomposition.  Explicit formulas will be given below.

\subsection{Conformal blocks from singlets}

In this subsection we describe how to holographically construct conformal blocks which can be combined to give crossing symmetric four-point functions of primary operators. We will focus on the holomorphic part of a conformal block denoted $g_p(z)$. This implies that we will ignore the representations $\tilde{\cal R}_i$ and deal only with the construction of singlets in the tensor product $\otimes_i {\cal R}_i.$

\begin{figure}[h]
  \centering
  \begin{minipage}[b]{0.48\textwidth}
    \centering\includegraphics[width=0.9\textwidth]{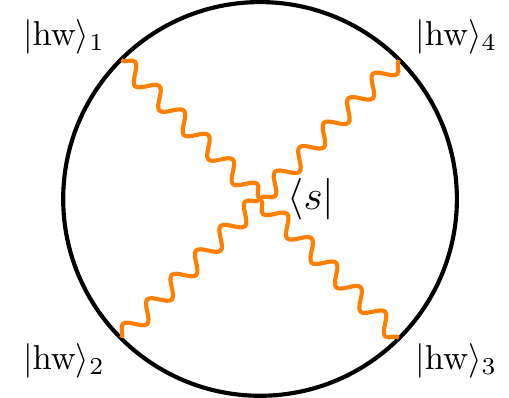}
    \caption*{a)}
  \end{minipage}
  \hfill
  \begin{minipage}[b]{0.48\textwidth}
   \centering\includegraphics[width=0.7\textwidth]{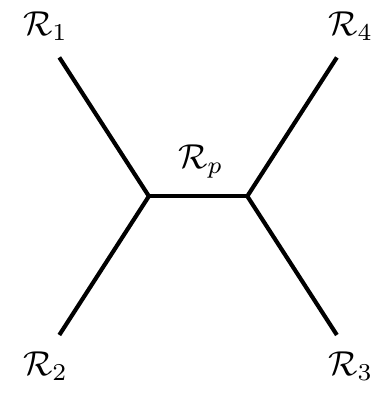}
    \caption*{b)}
  \end{minipage}
  \caption{a) Holographic calculation of a conformal block. Four bulk-to-boundary propagators consisting of Wilson lines in some representation ${\cal R}_i$ meet at a common bulk point where a singlet state is assembled. b) Construction of the singlet state $|s^{12,34}_p\rangle$. A representation ${\cal R}_p$ is chosen from the tensor product ${\cal R}_1\otimes {\cal R}_2$, while its conjugate $\overline{\cal R}_p$ is chosen from the tensor product ${\cal R}_3\otimes {\cal R}_4$. The singlet state is the one appearing in ${\cal R}_p\otimes \overline{\cal R}_p$. }
  \label{fig:f1}
\end{figure}

Following the discussion in section \ref{sec:GeneralFormulation} we consider four representations ${\cal R}_i$ of SL(N) and separate the operators into two pairs (12) and (34). These give rise to the tensor products
\begin{equation}
{\cal R}_1\otimes {\cal R}_2 =\oplus_a {\cal R}^{(12)}_a, \quad\quad  \quad\quad{\cal R}_3\otimes {\cal R}_4 =\oplus_a {\cal R}^{(34)}_a.
\end{equation}
Picking complex conjugate representations from the two sums we can construct singlets. We choose a representation ${\cal R}^{(12)}_a={\cal R}_p$ in the first sum, its conjugate ${\cal R}^{(34)}_a=\overline{\cal R}_p$ in the second sum and denote by $|s^{12,34}_p\rangle$ the singlet in ${\cal R}_p \otimes \overline{\cal R}_p$. Each singlet defines a conformal block when used in (\ref{ad}) which we adapt here to the case in consideration
\be
w_p(z_i) = \langle s^{12,34}_p| \prod_{i=1}^4  e^{z_{bi}T^{(i)}_1} |{\rm hw}\rangle_i~.
\ee
Figure \ref{fig:f1} shows a picture of this object.

Once we have obtained the blocks, the four point function can be constructed as
\be
G(x_i) = \langle {\cal O}_1(z_1,\zb_1) \ldots {\cal O}_4(z_4,\zb_4)\rangle = \sum_{p,\tilde{p}} A^{12,34}_{p\tilde{p}} w_p(z_i) \tilde{w}_{\tilde{p}}(\zb_i)
\ee
where $A^{12,34}_{p\tilde{p}}$ are in principle unknown constants related to the OPE coefficients as $A^{12,34}_{p\tilde{p}}=C^{12}_{p\tilde{p}} C^{34}_{p\tilde{p}}$. Alternatively, denoting the tensor product basis elements by $\vert S^{12,34}_{p\tilde{p}}\rangle \equiv |s^{12,34}_p\rangle |\tilde{s}^{12,34}_{\tilde{p}}\rangle$, we can define the singlet
\be\label{stzzbar}
|S\rangle = \sum_{p,\tilde{p}} A^{12,34}_{p\tilde{p}} |S^{12,34}_{p\tilde{p}}\rangle
\ee
and then write the four point function as
\be\label{fourzzbar}
G(z_i,\zb_i) = \langle S| \prod_{i=1}^4  e^{z_{bi}T^{(i)}_1} |{\rm hw}\rangle_i e^{\zb_{bi}\Tt^{(i)}_1} | \widetilde{\rm hw}\rangle_i~.
\ee

\subsection{Crossing symmetry}\label{sec:crossing}

In the above we expanded in the (12)(34) channel and wrote the corresponding basis of singlets as $\{|S^{12,34}_{p\tilde{p}}\rangle\}$, but we can expand in other channels as well, for example (14)(32). The corresponding basis of singlets will differ from the previous one and we denote it $\{|S^{14,32}_{p'\tilde{p}'}\rangle\}$. We can expand the singlet (\ref{stzzbar}) in the new basis
\be\label{sincross}
|S\rangle = \sum_{p,\tilde{p}} A^{12,34}_{p\tilde{p}} |S^{12,34}_{p\tilde{p}}\rangle = \sum_{p',\tilde{p}'} A^{14,32}_{p'\tilde{p}'} |S^{14,32}_{p'\tilde{p}'}\rangle.
\ee
The bases appearing in (\ref{sincross}) are complete and given coefficients $A^{12,34}_{p\tilde{p}}$ we can find coefficients $A^{14,32}_{p'\tilde{p}'}$ such that (\ref{sincross}) is obeyed. Crossing symmetry in the case that all ${\cal R}_i$ are distinct relates OPE coefficients in one channel to those of another.   The set of operators that appears in each channel has already been fixed by the rules above.

The situation changes if two of the operators  carry the same representation; for example suppose ${\cal R}_2={\cal R}_4$.   Then $G(x_i)$ should be invariant under $x_2 \leftrightarrow x_4$.  Looking at (\ref{fourzzbar}), this implies that $|S\rangle$ should be invariant under interchanging the states associated with ${\cal R}_2$ and ${\cal R}_4$.  This crossing symmetry condition imposes a constraint on the OPE coefficients. To see this we study the holomorphic singlet states $|s^{12,34}_p\rangle$.

The change of basis associated with $x_2 \leftrightarrow x_4$ is given by
\be
|s^{12,34}_{p}\rangle = \sum_{p'}O_{pp'}|s^{14,32}_{p'}\rangle
\ee
for some orthogonal matrix $O_{pp'}$ which we call the exchange matrix. We then have
\begin{align}
\begin{split}
|S\rangle &= \sum_{p,\tilde{p}} A^{12,34}_{p\tilde{p}} |s^{12,34}_p\rangle |\tilde{s}^{12,34}_{\tilde{p}}\rangle \\
&=\sum_{p,\tilde{p}} (O^{-1} A^{12,34}O)_{p\tilde{p}} |s^{14,32}_p\rangle |\tilde{s}^{14,32}_{\tilde{p}}\rangle~,
\end{split}
\end{align}
which implies $A^{14,32}=O^{-1} A^{12,34}O$.  This constraint on the OPE coefficients will play a role in section \ref{sec:slN} when we build four-point functions as sums over SL(N) conformal blocks.


%
%

\section{General SL(2) result}
\label{SL2}

We turn now to the evaluation of conformal blocks for the case of SL(2) representations. Each operator is associated with the highest weight state of a finite dimensional representation of SL(2). The Young tableaux for the representations ${\cal R}_i$ consist of a single row whose length is the Dynkin label $\lambda$.
\begin{align}
\begin{split}
{\cal R}_i=\underbrace{\Yboxdim{20pt}\young(\quad\quad
\Smalldots\quad\quad)}_{\lambda_i}=\{\lambda_i \} ~.
\end{split}
\end{align}
The Dynkin label is related to the spin of the representation as $\lambda_i=2j_i$. The conformal dimension associated to the highest weight state $\vert \text{hw}\rangle_i$ is given by $h_i=-\lambda_i/2=-j_i$. The negative value of $h$ is a manifestation of the non-unitary nature of the theory in which the primaries lie in finite dimensional representations of SL(2).   This will not pose any obstacle towards verifying precise and detailed agreement between bulk
and boundary observables in the limit of large central charge.

In this section we examine the calculation of a holographic conformal block whose external primary operators are highest weight states of representations ${\cal R}_i$ with Dynkin labels $\lambda_i$ placed at the points $z_i$ on the plane. Likewise, the exchanged primary is associated to a representation ${\cal R}_p$ with Dynkin label $\lambda_p$.  As explained above (\ref{ad}), the object we need to evaluate reads
\begin{align}\label{eq:gssl2}
\begin{split}
w_s (z_i)=\langle s \vert \prod_i e^{z_{bi}T^{(i)}_1}\vert \text{hw}\rangle_i~,
\end{split}
\end{align}
 where $\vert s\rangle$ is the singlet state corresponding to the exchange of the representation ${\cal R}_p$. Figure \ref{fig:f1}  shows an intuitive picture of the setup. We will implement the following strategy. First, we will construct the states of the representation ${\cal R}_p$ out of the states of ${\cal R}_1$ and ${\cal R}_2$. Likewise, we will obtain the states of $\overline{\cal R}_p$ out of those of ${\cal R}_3$ and ${\cal R}_4$. The singlet $\vert s \rangle$ is built by contracting all the SL(2) indices of the states in ${\cal R}_p$ with those of $\overline{\cal R}_p$ using the Levi-Civita symbol, which is an invariant tensor. To make the calculation easier, we will perform certain tricks involving gauge invariance. First, we will exploit conformal invariance to move three of the external primaries to $z_1=\infty$, $z_2=1$, and $z_3=0$. After this, the configuration of external primaries reads
\begin{align}
\begin{split}
 z_1=\infty\quad \text{:}&\quad  {\cal R}_1=\{\lambda_1 \} ~, \quad \quad  z_2=1\quad \text{:}\quad  {\cal R}_2=\{\lambda_2 \} ~,\\
 z_3=0\quad \text{:}&\quad  {\cal R}_3=\{\lambda_3 \} ~, \quad \quad  z_4=z\quad \text{:}\quad  {\cal R}_4=\{\lambda_4 \} ~.
\end{split}
\end{align}
Before attempting to write the singlet state $\vert s \rangle$, it is useful to notice that the Wilson line operator coming from infinity projects the highest weight state $\vert \text{hw}\rangle_1$  to the lowest weight state
\begin{align}\label{lowest weight}
\begin{split}
\lim_{z_1\rightarrow \infty}z^{2h_1}_1 e^{z_{b\infty}T^{(1)}_1}\vert \text{hw}\rangle_1  \propto \ket{-\text{hw}}_1 ~.
\end{split}
\end{align}
This observation simplifies the calculation of the singlet greatly, as we now need to focus only on the terms in $\vert s\rangle$ that are lowest weight for the primary ${\cal O}_1$. A further simplification of the calculation consists in choosing the bulk point where the Wilson lines meet to be at $z_b=0$. This gauge choice immediately implies that the Wilson line operator coming from the boundary point $z_3=0$ corresponds to the identity, and so it projects the highest weight state to itself.
 \begin{align}
\begin{split}
\lim_{z_b\rightarrow 0}e^{z_{b0}T^{(3)}_1}\vert \text{hw}\rangle_3 = \vert \text{hw}\rangle_3 ~.
\end{split}
\end{align}
As a consequence the only terms in $\vert s\rangle$ contributing to $w_s(z_i)$ are highest weight for ${\cal O}_3$ and lowest weight for ${\cal O}_1$. Instead of writing down $\langle s \vert$ we will compute $w_s(z_i)$ directly by replacing the states $\vert e_j\rangle_i$ by the objects $\qi_j \equiv \langle e_j \vert e^{z_{bi}T^{(i)}_1}\vert e_1 \rangle_i$, where $\vert e_j\rangle_i$ are the states in the defining representation of SL(2) and the subscript $i$ refers to the representation ${\cal R}_i$ (see appendices \ref{appthreepoint} and \ref{slnconventions}). We start with the following expressions for the Wilson line matrix elements involving states of the boundary representations
 \begin{align}
\begin{split}
\langle ({\cal R}_1)_{i_1\dots i_{\lambda_1}} \vert e^{z_{b1}T^{(1)}_1}\vert \text{hw} \rangle_1 &=  \delta^{2}_{i_1} \dots \delta^2_{i_{\lambda_1}} ~, \\
\langle ({\cal R}_2)_{i_1\dots i_{\lambda_2}} \vert e^{z_{b2}T^{(2)}_1}\vert \text{hw} \rangle_2 &=\qt_{(i_1} \dots \qt_{i_{\lambda_2})}  ~,\\
\langle ({\cal R}_3)_{i_1\dots i_{\lambda_3}} \vert e^{z_{b3}T^{(3)}_1}\vert \text{hw} \rangle_3 &=\delta^{1}_{i_1} \dots \delta^1_{i_{\lambda_3}} ~, \\
\langle ({\cal R}_4)_{i_1\dots i_{\lambda_4}} \vert e^{z_{b4}T^{(4)}_1}\vert \text{hw} \rangle_4 &=\qf_{(i_1} \dots \qf_{i_{\lambda_4})}  ~,
\end{split}
\end{align}
where we have projected the states of ${\cal R}_1$ to their lowest weight, and the states of ${\cal R}_3$ to their highest weight. We now build the representation ${\cal R}_p$ out of the states in the first pair. This representation must consist of $\lambda_p$ symmetric indices. There are a total of $\lambda_1+\lambda_2$ indices and each contraction with the Levi-Civita symbol subtracts two indices. It follows that     $(\lambda_1+\lambda_2-\lambda_p)/2$ contractions are needed. The result reads
\begin{align}\label{eq:Rpsl2}
\begin{split}
\langle ({\cal R}_p)_{i_1\dots i_{\lambda_p}} \vert e^{z_{b1}T^1_1}\vert \text{hw} \rangle_1 e^{z_{b2}T^2_1}\vert \text{hw} \rangle_2&=(\qt_1)^{{{\lambda_1+\lambda_2-\lambda_p}\over 2}} \delta^2_{(i_1}\dots\delta^2_{i_{{{\lambda_p+\lambda_1-\lambda_2}\over 2}}}\qt_{i_{{{\lambda_p+\lambda_1-\lambda_2}\over 2}+1}}\dots \qt_{i_{\lambda_p})}~.
\end{split}
\end{align}
The same logic follows for the construction of the states in $\overline{\cal R}_p$. In this case, there will be $(\lambda_3+\lambda_4-\lambda_p)/2$ contractions with the Levi-Civita symbol
\begin{align}\label{eq:Rpbsl2}
\begin{split}
\langle (\overline{\cal R}_p)_{i_1\dots i_{\lambda_p}} \vert e^{z_{b3}T^3_1}\vert \text{hw} \rangle_3 e^{z_{b4}T^4_1}\vert \text{hw} \rangle_4&=(\qf_2)^{{{\lambda_3+\lambda_4-\lambda_p}\over 2}} \delta^1_{(i_1}\dots\delta^1_{i_{{{\lambda_p+\lambda_3-\lambda_4}\over 2}}}\qf_{i_{{{\lambda_p+\lambda_3-\lambda_4}\over 2}+1}}\dots \qf_{i_{\lambda_p})}~.
\end{split}
\end{align}
Finally, the singlet is obtained by contracting all the indices of (\ref{eq:Rpsl2}) with the indices of (\ref{eq:Rpbsl2}) using Levi-Civita symbols:
\begin{align}\label{eq:gssl2sum}
\begin{split}
g_s(z)&=(\qt_1)^{{{\lambda_1+\lambda_2-\lambda_p}\over 2}}(\qf_2)^{{{\lambda_3+\lambda_4-\lambda_p}\over 2}}\epsilon^{i_1 j_1}\dots\epsilon^{i_{\lambda_p} j_{\lambda_p}} \\
&\times  \delta^2_{(i_1}\dots\delta^2_{i_{{{\lambda_p+\lambda_1-\lambda_2}\over 2}}}\qt_{i_{{{\lambda_p+\lambda_1-\lambda_2}\over 2}+1}}\dots \qt_{i_{\lambda_p})}\times  \delta^1_{j_1}\dots\delta^1_{j_{{{\lambda_p+\lambda_3-\lambda_4}\over 2}}}\qf_{j_{{{\lambda_p+\lambda_3-\lambda_4}\over 2}+1}}\dots \qf_{j_{\lambda_p}} ~.
\end{split}
\end{align}
The last step of the calculation is to evaluate the object (\ref{eq:gssl2sum}). The strategy is the following: we first classify the different symmetric permutations that give rise to inequivalent contributions to $g_s(z)$. We will then sum over all permutation classes, taking into account their contribution and multiplicity.

In order to classify the different permutations, let us define ``red"  indices as the indices appearing in the objects $\delta^1_j$. We also define ``green" indices as the indices appearing in $\qf_j$. Each permutation will contribute differently depending of how many red and green indices appear in the delta functions $\delta^2_i$ (Box 1) and the objects $\qt_i$ (Box 2). We then define our permutation class as those with $k$ red indices in Box 1. This also implies there will be ${{\lambda_p+\lambda_1-\lambda_2}\over 2 }-k$ green indices in Box 1, ${{\lambda_p+\lambda_3-\lambda_4}\over 2}-k$ red indices in Box 2, and ${{\lambda_4-\lambda_3+\lambda_2-\lambda_1}\over 2}+k$ green indices in Box 2. Each permutation of this class will contribute to the block as follows
\begin{align}\label{classk}
\begin{split}
 g^{(k)}_s(z)&=(\qt_1)^{{{\lambda_1+\lambda_2-\lambda_p}\over 2}}(\qf_2)^{{{\lambda_3+\lambda_4-\lambda_p}\over 2}}\\
&\times (\epsilon^{i_R j_R}\delta^2_{i_R}\delta^1_{j_R})^{k}  (\epsilon^{i_G j_G} \delta^2_{i_G}\qf_{j_G})^{{{\lambda_p+\lambda_3-\lambda_4}\over 2}-k} (\epsilon^{i_R j_R }\qt_{i_R} \delta^{1}_{j_R})^{{{\lambda_p+\lambda_3-\lambda_4}\over 2}-k}  (\epsilon^{i_G j_G} \qt_{i_G}\qf_{j_G})^{{{\lambda_4-\lambda_3+\lambda_2-\lambda_1}\over 2}+k}\\
&=z^{{{\lambda_3+\lambda_4-\lambda_p}\over 2}} (1-z)^{{{\lambda_4-\lambda_3+\lambda_2-\lambda_1}\over 2}+k}~.
\end{split}
\end{align}
The multiplicity of each class consists of choosing $k$ red indices out of a total of ${{\lambda_p+\lambda_3-\lambda_4}\over 2}$, choosing ${{\lambda_p+\lambda_1-\lambda_2}\over 2 }-k$ green indices out of a total of ${{\lambda_p+\lambda_4-\lambda_3}\over 2}$, and ordering the indices of each box. We then have
\begin{align}\label{classk1}
\begin{split}
C^{(k)}={{{{\lambda_p+\lambda_3-\lambda_4}\over 2}}\choose{k}} {{{{\lambda_p+\lambda_4-\lambda_3}\over 2}}\choose {{{\lambda_p+\lambda_1-\lambda_2}\over 2 }-k}} \Gamma \left({{\lambda_p+\lambda_1-\lambda_2}\over 2}+1\right) \Gamma \left({{\lambda_p+\lambda_4-\lambda_3}\over 2}+1\right) ~.
\end{split}
\end{align}
We are now ready to sum over permutation classes. This consists of a sum over $k$. The result reads
\begin{align}\label{classk2}
\begin{split}
 g_s(z)&= \sum^{{{\lambda_p+\lambda_3-\lambda_4}\over 2}}_{k=0} C^{(k)} g^{(k)}_s(z)=z^{{{\lambda_3+\lambda_4-\lambda_p}\over 2}} {_2 F}_1\left(-{{\lambda_p+\lambda_2-\lambda_1}\over 2},-{{\lambda_p+\lambda_4-\lambda_3}\over 2};-\lambda_p;z\right)~.
\end{split}
\end{align}
This result can be written in a more suggestive way by replacing $\lambda_i\rightarrow -2 h_i$
\begin{align}
\begin{split}
g_s(z)&=z^{-h_3-h_4+h_p} {_2F}_1\left(h_p+h_{21},h_p+h_{43};2h_p;z\right)~.
\end{split}
\end{align}
This is the standard result for the chiral half of the global conformal block \cite{Dolan:2011dv}.  This result was also obtained in \cite{Bhatta:2016hpz}.

\section{SL(3) Result}\label{sl3}

After the warmup with \SL{2}, we can now move on to the more difficult task of computing \SL{3} blocks. Our goal here is to compute conformal blocks of ${\cal W}_3$ in the large central charge limit with the operator dimensions and charges kept fixed as $c\to \infty$. The ${\cal W}_3$ algebra reduces to SL(3) in the large central charge limit. Our strategy as before will be to compute blocks in finite dimensional representations of \SL{3} and then continue the result to more general representations. Finite dimensional irreducible representations of SL(3) are labelled by two integers (the Dynkin labels) $\lo$ and $\lt$. Alternatively, they can be written as symmetric traceless tensors with $\lo$ lower and $\lt$ upper indices where the lower and upper indices denote states in the defining representation and its conjugate respectively (see appendix \ref{group theory} for details). Our main goal in this section is to reproduce the result for ${\cal W}_3$ conformal blocks obtained in \cite{Fateev:2011qa}. 

In terms of \SL{3} tensors, constructing the singlet amounts to contracting all lower and upper indices. It turns out to be computationally more tractable if we consider two of the representations to have only upper (or only lower) indices i.e. the tensor product $(\lo,\lt)\otimes(0,\mu)\otimes(0,\mup)\otimes(\lop,\ltp)$. Let the exchanged representation be $\Rc_p=(x,y)$. Below we list the Young tableaux associated to these representations
\al{f}{
\Yvcentermath1
\begin{array}{clcclc}
\Rc_1\quad & \Yboxdim{20pt}\young(\lt\lo,\lt)&\quad z_1=0~,&\quad\Rc_2 \quad & \Yboxdim{20pt}\young(\mu,\mu)&\quad z_2=z~,
\\
&&&&&\\
\Rc_3\quad & \Yboxdim{20pt}\young(\mup,\mup)&\quad z_3=1~,&\quad\Rc_4 \quad & \Yboxdim{20pt}\young(\ltp\lop,\ltp)&\quad z_4=\infty~,
\\
&&&&\\
\Rc_p\quad & \Yboxdim{20pt}\young(yx,y)&\quad z_p=z_b&&&
\end{array}
}
To avoid cluttering, in the above Young tableau we have used $\Yboxdim{15pt}\young(\lambda)$ to denote a row of $\lambda$ boxes. To evaluate the conformal block in the $(12)(34)$ channel, we first construct the tensor products $\Rc_1\otimes\Rc_2$ and $\Rc_3\otimes\Rc_4$ in terms of \SL{3} tensors. The singlet is then obtained by contracting all indices between the tensors coming from the two tensor products\footnote{For a singlet to exist, the two irreps coming from the two tensor products must be conjugate to each other. Since conjugating irreps of \SL{3} is equivalent to switching the Dynkin labels, the singlet exists only if the number of upper indices on the first tensor is equal to the number of lower indices on the second and vice versa.}. To be a little more explicit, the representation $\Rc_p$ in the tensor product $\Rc_1\otimes\Rc_2$ can be written as
\al{fa}{
M_{i_1\cdots i_x}^{j_1\cdots j_y} = (P_{i_1\cdots i_x}^{j_1\cdots j_y})_{b_1\cdots b_{\lt} c_1\cdots c_\mu}^{a_1\cdots a_{\lo}}\ket{e_{a_1}\ldots e_{a_{\lo}}\bar{e}^{b_1}\ldots \bar{e}^{b_{\lt}}\bar{e}^{c_1}\ldots \bar{e}^{c_{\mu}}}
}
where the indices $a$ and $b$ denote states of the representation $\Rc_1$ and $c$ of $\Rc_2$. As a consequence all the $a$, $b$ and $c$ indices are symmetrized and any contraction between $a$ and $b$ vanishes. The tensor $P$ projects onto the representation $(x,y)$ and as we explain below must be built out of $\delta_k^l$'s and $\epsilon_{klm}$'s. Note that for the new tensor $M$ to be irreducible, it must be completely symmetric and traceless. The tensor $N$ for the representation $\overline{\Rc}_p$ can be constructed out of the tensor product $\Rc_3\otimes\Rc_4$ in a similar manner.
\al{faa}{
N^{i_1\cdots i_x}_{j_1\cdots j_y} = (P^{i_1\cdots i_x}_{j_1\cdots j_y})_{g_1\cdots g_{\ltp} h_1\cdots h_{\mup}}^{f_1\cdots f_{\lop}}\ket{\bar{e}^{h_1}\ldots \bar{e}^{h_{\mup}}e_{f_1}\ldots e_{f_{\lop}}\bar{e}^{g_1}\ldots \bar{e}^{g_{\ltp}}}
}
where the indices $f$ and $g$ denote states of $\Rc_4$ and $h$ of $\Rc_3$. In the full tensor product the singlet state is then obtained as
\al{fb}{
\ket{s} = M_{i_1\cdots i_x}^{j_1\cdots j_y}N^{i_1\cdots i_x}_{j_1\cdots j_y}
}
Now let's study the kinds of representations that can appear in \eqref{fa}. We start out with $\lo$ lower indices and $\lt+\mu$ upper indices. The operations we can perform that are invariant under \SL{3} are contraction with the invariant tensors $\delta_k^l$, $\epsilon_{klm}$ and $\epsilon^{klm}$. Taking the symmetry properties of $a$, $b$ and $c$ into account, we are allowed to do one of two things: contract indices $a$ and $c$ using $\delta^{a}_{c}$, or convert indices $b$ and $c$ into a lower index using $\epsilon_{ibc}$. If we perform $d$ contractions using $\delta$'s and $e$ conversions using $\epsilon$'s, a simple counting of indices requires the relation
\al{fc}{
(x,y) = (\lo-d+e,\lt+\mu-d-2e)
}
We still need to make the tensor symmetric and traceless. The procedure for making a symmetric tensor traceless is described in appendix \ref{traceless}. We will deal with this later as it doesn't change the relation in \eqref{fc}. Performing similar operations on the $\Rc_3\otimes\Rc_4$ tensor product with $d'$ contractions and $e'$ conversions, we obtain
\al{fd}{
(y,x) = (\lop-d'+e',\ltp+\mup-d'-2e')
}
The projectors in \eqref{fa} and \eqref{faa} without the tracelessness constraint imposed now look like
\al{fe}{
(P_{i_1\cdots i_x}^{j_1\cdots j_y})_{b_1\cdots b_{\lt} c_1\cdots c_\mu}^{a_1\cdots a_{\lo}} &= \delta_{i_1}^{a_1}\cdots\delta_{i_{l_1}}^{a_{l_1}}\delta_{b_1}^{j_1}\cdots\delta_{b_{l_3}}^{j_{l_3}}\delta_{c_1}^{j_{l_3+1}}\cdots\delta_{c_{l_4}}^{j_{y}}\epsilon_{i_{l_1+1}b_{l_3+1}c_{l_4+1}}\cdots\epsilon_{i_x b_{\lt}c_{l_4+l_2}}
\\
&\quad\times \delta^{a_{l_1+1}}_{c_{l_4+l_2+1}}\cdots\delta^{a_{\lo}}_{c_{\mu}}
\\
(P^{i_1\cdots i_x}_{j_1\cdots j_y})_{g_1\cdots g_{\ltp} h_1\cdots h_{\mup}}^{f_1\cdots f_{\lop}} &= \delta_{j_1}^{f_1}\cdots\delta_{j_{n_4}}^{f_{n_4}}\delta_{g_1}^{i_1}\cdots\delta_{g_{n_1}}^{i_{n_1}}\delta_{h_1}^{i_{n_1+1}}\cdots\delta_{h_{n_2}}^{i_{x}}\epsilon_{j_{n_4+1}g_{n_1+1}h_{n_2+1}}\cdots\epsilon_{j_y g_{n_1+n_5}h_{n_2+n_5}}
\\
&\quad\times \delta^{f_{n_4+1}}_{h_{n_2+n_5+1}}\cdots\delta^{f_{\lop}}_{h_{\mup}}
}
where the $i$'s and $j$'s are to be completely symmetrized. We have made the following definitions for notational convenience
\al{ff}{
&l_1=\lo-d~,~~l_2=e~,~~l_3=\lt-e~,~~l_4=\mu-d-e
\\
&n_1=\ltp-e'~,~~n_2=\mup-d'-e'~,~~n_4=\lop-d'~,~~n_5=e'~,~~
}
In simple terms, $l_1$ is the number of $i$ indices that appear in $\delta^a_i$, the rest of them ($l_2$ in number) being in $\epsilon_{ibc}$. Similarly, $n_1$ is the number of $j$ indices that appear in $\delta_b^j$, the rest of them ($n_2$ in number) being in $\delta_c^j$ and so on. Recall that our final goal is to calculate the Wilson line
\al{fg}{
w_s(z_k) = \langle s| \prod_{k=1}^4  e^{z_{bk}T^{(k)}_1} |{\rm hw}\rangle_k
}
To make direct comparison with the results of \cite{Fateev:2011qa}, we choose the operator positions -- $z_1=0,z_2=z,z_3=1,z_4=\infty$ where $z_k$ denotes the position associated to the representation $\Rc_k$. To this end we define
\al{fh}{
\qk_a=\braket{e_a|e^{z_{bk}T^{(k)}_1}|e_1}_k~,~~\qbk^{b} = \braket{\bar{e}^b|e^{z_{bk}T^{(k)}_1}|\bar{e}^3}_k
}
using which we can directly write out the matrix elements rather than the states appearing in the singlet $\ket{s}$. The contributions from the $\Rc_1\otimes \Rc_2$ tensor product can then be written as
\al{fi}{
(M_{0z})_{i_1\cdots i_x}^{j_1\cdots j_y} &= \left((M^\dagger)_{i_1\cdots i_x}^{j_1\cdots j_y}\right)e^{z_{b1}T^{(1)}_1}\otimes e^{z_{b2}T^{(2)}_1}|{\rm hw}\rangle_1|{\rm hw}\rangle_2
\\
&= \qo_{(i_1}\cdots\qo_{i_{l_1}}\tq_{i_{l_1+1}}\cdots\tq_{i_x)}\qbo^{(j_1}\cdots\qbo^{j_{l_3}}\qbt^{j_{l_3+1}}\cdots\qbt^{j_{y})}
\\
&\quad\times \delta^{a_{l_1+1}}_{c_{l_4+l_2+1}}\cdots\delta^{a_{\lo}}_{c_{\mu}}\qo_{a_{l_1+1}}\qbt^{c_{l_4+l_2+1}}\cdots\qo_{a_{\lo}}\qbt^{c_{\mu}}
}
with $\tq_i \equiv \epsilon_{ibc}\qbo^b\qbt^c$. In the tensor $M_{0z}$, it is clear that there are two types of lower indices: ones that appear on $\qo$ and ones on $\tq$. There are two types of upper indices too: ones on $\qbo$ and ones on $\qbt$. As in the \SL{2} case, we refer to these different types of indices by colors. The indices on $\qo$ we call red, $\tq$ blue, $\qbo$ green and $\qbt$ yellow. In this language, the $l_i$ defined in \eqref{ff} are just the number of indices of each color. There are further simplifications once we fix the positions of the bulk and boundary points. We use our freedom of choosing the bulk point to set $z_b=0$ such that the Wilson line projects out the highest weight state of $\Rc_1$. This forces all red and green indices to be highest weight indices i.e. $\qo_1$ and $\qbo^3$ respectively.

A similar story plays out for the $\Rc_3\otimes \Rc_4$ tensor product
\al{fj}{
(N_{1\infty})^{i_1\cdots i_x}_{j_1\cdots j_y} &= \qf_{j_1}\cdots\qf_{j_{n_4}}\tq'_{j_{n_4+1}}\cdots\tq'_{j_y}\qbf^{i_1}\cdots\qbf^{i_{n_1}}\qbth^{i_{n_1+1}}\cdots\qbth^{i_{x}}
\\
&\quad\times \delta^{f_{n_4+1}}_{h_{n_2+n_5+1}}\cdots\delta^{f_{\lop}}_{h_{\mup}}\qf_{f_{n_4+1}}\qbth^{h_{n_2+n_5+1}}\cdots\qf_{f_{\lop}}\qbth^{h_{\mup}}
}
with $\tq'_{j}=\epsilon_{jgh}\qbf^{g}\qbth^{h}$. Again mimicking the \SL{2} computations, we refer to the indices of $N_{1\infty}$ as boxes. We call the indices on $\qbf$ and $\qbth$ box 1 and box 2 respectively. Saving box 3 for a different purpose, we call the indices on $\qf$ and $\tq'$ box 4 and box 5 respectively. The $n_i$ of \eqref{ff} count the number of boxes of each type. As discussed before in \eqref{lowest weight}, setting $z_4\to\infty$ projects out the lowest weight state from the singlet in the Wilson line. In other words all box 1 and box 4 indices are forced to be lowest weight indices i.e. $\qbf^1$ and $\qf_3$ respectively.

The explicit form of the matrices $L_1$ and $L_{-1}$ (see appendix \ref{group theory}) in the defining representation gives
\al{fk}{
\qk_a &= \braket{e_a|e^{-z_k\lp}|e_1}_k = \delta^1_a + \sqrt{2}z_k\delta^2_a + z_k^2\delta^3_a
\\
\qbk^a &= \braket{\bar{e}^a|e^{-z_k\lm}|\bar{e}^3}_k = \delta_3^a - \sqrt{2}z_k\delta_2^a + z_k^2\delta_1^a
}
Using $z_1 = 0$, $z_2=z$ and the fact that all the $\qo$ indices are highest weight indices we have $\delta^a_c\qo_a\qbt^c = \delta^1_c\qo_1\qbt^c = z^2$. Similarly, all $\qf$ indices are lowest weight giving $\delta^{f}_{h}\qf_{f}\qbth^{h} = \delta^{3}_{h}\qf_{3}\qbth^{h} = 1$.

Next , let us deal with the issue of making the exchanged tensor traceless. As discussed in appendix \ref{traceless}, we first subtract all possible traces of the tensor. Then we subtract out traces of the new terms added and so on until we run out of traces. The result from \eqref{a3j} is
\al{fl}{
(\tilde{N}_{1\infty})^{i_1\cdots i_x}_{j_1\cdots j_y} = \sum_{n=0}^{min(x,y)}C_n\delta^{(i_1}_{(j_1}\cdots\delta^{i_n}_{j_n}(N_{1\infty})^{i_{n+1}\cdots i_x)k_1\cdots k_n}_{j_{n+1}\cdots j_y)k_1\cdots k_n}
}
where the $C_n$ are read off from \eqref{a3j}. In doing this we have introduced new types of upper and lower indices -- the ones appearing on $\delta^i_j$. We call the upper index box 3, and lower box 6. A caveat here is that the trace of some indices vanishes like the ones coming from the representation $(\lop,\ltp)$. In other words some terms in the symmetrization in \eqref{fl} vanish depending on what indices are being traced out. Note that in the absence of this constraint all terms in the symmetrization would contribute in exactly the same manner. To account for the constraint we simply assume that all possible traces are allowed but then correct by multiplying by the fraction of terms that would survive in the symmetrization. From \eqref{fj}, we see $\qbf\cdot\tq' = 0 = \qbth\cdot\tq' = \qbf\cdot\qf$ allowing us to trace out only box 2 and box 4. The fraction of terms for a given value of $n$ is then found as follows : choose $n$ indices from box 2 and box 4 to trace out, multiply by the number of permutations that preserve this structure and divide by the total number of terms. This gives an additional factor to add onto \eqref{fl}
\al{fm}{
C'_n &= {n_2\choose n}{n_4\choose n}\frac{\Gamma(x-n+1)\Gamma(y-n+1)\Gamma(n+1)^2}{\Gamma(x+1)\Gamma(y+1)}
\\
&=\frac{\Gamma(n_2+1)\Gamma(n_4+1)\Gamma(x-n+1)\Gamma(y-n+1)}{\Gamma(n_4-n+1)\Gamma(n_2-n+1)\Gamma(x+1)\Gamma(y+1)}
}
We now have the two objects $\tilde{M}_{0z}$ and $\tilde{N}_{1\infty}$ and the only thing left to do is to contract the indices between them in order to assemble the singlet. Note that since we are contracting all indices, it is sufficient to make just one of them symmetric and traceless. Putting all of this together, we have
\al{fn}{
g_s(z) &= (M_{0z})_{i_1\cdots i_x}^{j_1\cdots j_y}(\tilde{N}_{1\infty})^{i_1\cdots i_x}_{j_1\cdots j_y}
\\&= \qo_{i_1}\cdots\qo_{i_{l_1}}\tq_{i_{l_1+1}}\cdots\tq_{i_x}\qbo^{j_1}\cdots\qbo^{j_{l_3}}\qbt^{j_{l_3+1}}\cdots\qbt^{j_{y}}\times z^{2d}
\\
&~\times\sum_{n=0}^{min(x,y)}C_nC'_n\delta^{(i_1}_{(j_1}\cdots\delta^{i_n}_{j_n}\qf_{j_{n+1}}\cdots\qf_{j_{n_4}}\tq'_{j_{n_4+1}}\cdots\tq'_{j_y}\qbf^{i_{n+1}}\cdots\qbf^{i_{n_1+n+1}}\qbth^{i_{n_1+n+2}}\cdots\qbth^{i_{x}}\times 1
}
As for the \SL{2} case, keeping track of the permutations is a combinatorial problem; we need to find different ways to color boxes 1, 2 and 3 red or blue and boxes 4, 5 and 6 green or yellow. The details are relegated to appendix \ref{combinatorics}. Ignoring all factors that are independent of $z$ and the integer $n$, we obtain
\al{fo}{
g_s(z) = z^{2d}\sum_{n=0}^{\infty}&\frac{z^{2n}}{n!}\frac{(-n_2)_n(-l_4)_n(-l_1)_n(-n_4)_n}{(-x)_n(-y)_n(-x-y-1)_n}
\\
&\times {_2F}_1(-l_2,n-n_2;n-x;z){_2F}_1(-n_5,n-l_4;n-y;z)
}
The representations we consider here are of the same form as the ones in \cite{Fateev:2011qa}. The $r_i$ and $s_i$ there are defined to be the negative of the Dynkin labels : $r_1 = -\lo$, $s_1=-\lt$ and so on. Using this we find the following map to the definitions in equation (2.65) of \cite{Fateev:2011qa} : $n_5\to-\alpha$, $l_2\to -\beta$, $n_2\to -\gamma$ and $\l_4\to -\delta$. With these relations our result in \eqref{fo} agrees with their CFT calculation of the ${\cal W}_3$ blocks in the large $c$ limit.

\section{ An SL(N) example}\label{sec:slN}

We now consider an example at arbitrary $N$, but with simple representations so as to keep the computation tractable.
In particular, we will study the four-point function of two primaries in the fundamental (defining) representation of SL(N), and two primaries in the anti-fundamental representation of SL(N),
\begin{equation}\label{pprxi}
\langle{\cal O}_1(x_1){\cal O}_2(x_2){\cal O}_3(x_3){\cal O}_4(x_4) \rangle=\langle \phi_+(x_1)\overline{\phi}_+(x_2)\phi_+(x_3)\overline{\phi}_+(x_4) \rangle ~.
\end{equation}
Using conformal invariance and identifying the conformal cross ratios
\begin{align}\label{ccross}
z=\frac{z_{12}z_{34}}{z_{13}z_{24}}~,\quad \zb=\frac{\zb_{12}\zb_{34}}{\zb_{13}\zb_{24}}
\end{align}
this reduces to
\begin{equation}\label{ppr}
G_{\phi_+\overline{\phi}_+\phi_+\overline{\phi}_+}=\langle \phi_+(\infty)\overline{\phi}_+(1,1)\phi_+(z,\zb)\overline{\phi}_+(0,0) \rangle ~,
\end{equation}
where ${\cal O}_1={\cal O}_2=\phi_+$ and ${\cal O}_3={\cal O}_4=\overline{\phi}_+$ are primaries corresponding to the highest weight states of the following representations
\begin{align}
\begin{split}
\phi_+ \quad \text{:} \quad {\cal R}_1={\cal R}_2={\cal R}_+=\left(\yng(1),0\right) \quad \quad\text{and} \quad \quad \overline{\phi}_+\quad \text{:}\quad \overline{{\cal R}}_3=\overline{{\cal R}}_4=\overline{{\cal R}}_+=\left(\overline{\yng(1)},0\right)~.
\end{split}
\end{align}
We denote by $\vert \text{hw}\rangle_{i}$ the highest weight state of ${\cal R}_i$, and by $\vert \overline{\text{hw}}\rangle_{i}$ the highest weight state of $\overline{\cal R}_i$.

The holographic calculation of the blocks corresponding to this four point function follows the logic of section \ref{sec:4pt}. We first construct the matrix elements of the Wilson lines acting on the boundary states.  We then build the states corresponding to the exchanged representations, and we end the calculation by assembling the singlet. We will work in the channel where the pair $\phi_+(\infty)\overline{\phi}_+(1)$ exchanges states with the pair $\phi_+(z)\overline{\phi}_+(0)$. In order to see what representations can be exchanged, we decompose the tensor product of the representations of $\phi_+$ and $\overline{\phi}_+$
\begin{align}
\Yvcentermath1
\begin{split}
\yng(1)\otimes\overline{\yng(1)}=\mathbf{1}\oplus \mathbf{Adj}~, \quad \quad \text{where}\quad \quad  \mathbf{Adj}=\Yboxdim{20pt}\young(\quad\quad,\vdots,\quad)~.
\end{split}
\end{align}
The adjoint representation is conjugate to itself, so there are two different blocks we can construct. One of them corresponds to the exchange of the identity representation, the other corresponds to the exchange of the adjoint representation. We construct each block in a separate subsection.

\subsection{Exchange of $\mathbf{1}$}
We start by building the matrix elements of the bulk-to-boundary Wilson lines acting on the highest weight states at the boundary.
\begin{align}\label{RislN}
\begin{split}
\langle ({\cal R}_1)_{j} \vert e^{z_{b1}L_1} \vert \text{hw}\rangle_{1}=\qo_j ~,\\
\langle ({\cal R}_2)^{k} \vert e^{z_{b2}L_1} \vert \overline{\text{hw}}\rangle_{2}=\qbt^k~,\\
\langle ({\cal R}_3)_{j} \vert e^{z_{b3}L_1} \vert \text{hw}\rangle_{3}=\qth_j ~,\\
\langle ({\cal R}_4)^{k} \vert e^{z_{b4}L_1} \vert \overline{\text{hw}}\rangle_{4}=\qbf^k ~,
\end{split}
\end{align}
with $\qi_j=\langle e_j \vert e^{z_{bi}T^{(i)}_1} \vert \text{hw}\rangle_i$ and $\qbi^k=\langle e^{k}\vert e^{z_{bi}T^{(i)}_1}\vert \overline{\text{hw}}\rangle_i$.
The next step is to build the trivial representation out of each pair. We do this by contracting indices with the invariant tensor $\delta^{j}_{k}$
\begin{align}
\begin{split}
\langle (\mathbf{1}) \vert e^{z_{b1}L_1} \vert \text{hw}\rangle_{1}e^{z_{b2}L_1} \vert \overline{\text{hw}}\rangle_{2}=\qo_j \qbt^k\delta^{j}_{k}~,\\
\langle (\overline{\mathbf{1}}) \vert e^{z_{b3}L_1} \vert \text{hw}\rangle_{3}e^{z_{b4}L_1} \vert \overline{\text{hw}}\rangle_{4}=\qth_j \qbf^k\delta^{j}_{k}~.
\end{split}
\end{align}
The last step is to assemble the singlet out of $\mathbf{1}$ and $\overline{\mathbf{1}}$. No contractions with any tensor are needed
\begin{align}
\begin{split}
w_{\mathbf{1}}(z_i)=\langle s \vert e^{z_{b1}L_1} \vert \text{hw}\rangle_{1}e^{z_{b2}L_1} \vert \overline{\text{hw}}\rangle_{2}\vert e^{z_{b3}L_1} \vert \text{hw}\rangle_{3}e^{z_{b4}L_1} \vert \overline{\text{hw}}\rangle_{4}=\frac{1}{N}(\qo_j \qbt^k\delta^{j}_{k})( \qth_{j'} \qbf^{k'}\delta^{j'}_{k'})~
\end{split}
\end{align}
where we normalized the singlet. Using the explicit form of $\qi_j$ and $\qbi^{j}$ we obtain
\begin{align}
\begin{split}
g_{\mathbf{1}}(z)=\frac{1}{N}z^{N-1}~.
\end{split}
\end{align}

\subsection{Exchange of $\mathbf{Adj}$}
We proceed in the same fashion as in the previous subsection. We start with the expressions for the matrix elements of the bulk-to-boundary Wilson lines acting on the highest weight states at the boundary. These are written in (\ref{RislN}). The next step is to build the adjoint representation using the matrix elements of the first pair, the same can be done for the second pair. For this we need an object with one index down (index in the fundamental representation), and one index up (index in the anti-fundamental representation). For the representation to be irreducible we also must impose a tracelessness condition. The answer reads
\begin{align}
\begin{split}
(M_{12})^{k}_{\ j}\equiv \langle (\mathbf{Adj})^{k}_{\  j} \vert e^{z_{b1}L_1} \vert \text{hw}\rangle_{1}e^{z_{b2}L_1} \vert \overline{\text{hw}}\rangle_{2}=\qo_j \qbt^k-{1\over N}\delta^k_j\qo_i \qbt^i ~,\\
(M_{34})^{k}_{\ j}\equiv\langle (\overline{\mathbf{Adj}})^k_{\ j} \vert e^{z_{b3}L_1} \vert \text{hw}\rangle_{3}e^{z_{b4}L_1} \vert \overline{\text{hw}}\rangle_{4}=\qth_j\qbf^k -{1\over N}\delta^k_j\qth_i \qbf^i~,
\end{split}
\end{align}
where the second term in each expression ensures tracelessness. The singlet can now be built by contracting all indices of $ (M_{12})^{k}_{\ j}$ with all indices of $(M_{34})^k_{\ j}$ using Kronecker delta functions
\begin{align}
\begin{split}
w_{\mathbf{Adj}}(z_i) &= \langle s \vert e^{z_{b1}L_1} \vert \text{hw}\rangle_{1}e^{z_{b2}L_1} \vert \overline{\text{hw}}\rangle_{2}e^{z_{b3}L_1} \vert \text{hw}\rangle_{3}e^{z_{b4}L_1} \vert \overline{\text{hw}}\rangle_{4}\\
&=\frac{1}{\sqrt{N^2-1}}\delta^{j}_k\delta^{j'}_{k'}(M_{12})^{k'}_{\ j} (M_{34})^{k}_{\ j'} \\
&=\frac{1}{\sqrt{N^2-1}}\left( \qo_j \qbt^k \qth_k\qbf^j - {1\over N} \qo_j \qbt^j \qth_k\qbf^k \right)
~.
\end{split}
\end{align}
Using the explicit form of $\qi_j$ and $\qbi^{j}$ we obtain
\begin{align}
\begin{split}
g_{\mathbf{Adj}}(z)= \frac{1}{\sqrt{N^2-1}}\left( {(z-1)}^{N-1}-{1\over N}z^{N-1} \right)~.
\end{split}
\end{align}

\subsection{The four-point function $G_{\phi_+\overline{\phi}_+\phi_+\overline{\phi}_+}$}
As explained below (\ref{ac}), we have only computed the holomorphic conformal blocks. In order to obtain the four point function we need to sum over the products of holomorphic conformal blocks $g_s(z)$ and anti-holomorphic conformal blocks $\tilde{g}_{\tilde{s}}(\overline{z})$. We then write
\begin{align}\label{pprcb}
\begin{split}
G_{\phi_+\overline{\phi}_+\phi_+\overline{\phi}_+} =\sum_{p,\tilde{p}=\mathbf{1},\mathbf{Adj}} A_{p\tilde{p}} g_p(z)\tilde{g}_{\tilde{p}}(\overline{z})=g^T(z)Ag(\zb) ~,
\end{split}
\end{align}
where we introduced the matrix $A$ and the vectors
\begin{align}
g(z)=\begin{pmatrix} g_{\mathbf{1}}(z) \\ g_{\mathbf{Adj}}(z) \end{pmatrix}~,\quad g(\zb)=\begin{pmatrix} g_{\mathbf{1}}(\zb) \\ g_{\mathbf{Adj}}(\zb) \end{pmatrix}.
\end{align}
The correlator written explicitly in (\ref{ppr}) is invariant under the exchange $x_2 \leftrightarrow x_4.$ This translates to a  constraint on the matrix $A$ in our construction. To see this we first observe that the vector $g(z)$ transforms under the exchange as
\begin{align}
g(z)=\begin{pmatrix} g_{\mathbf{1}}(z) \\ g_{\mathbf{Adj}}(z) \end{pmatrix} ~\rightarrow~ \begin{pmatrix} \frac{1}{N} & \frac{\sqrt{N^2-1}}{N} \\ \frac{\sqrt{N^2-1}}{N} & -\frac{1}{N}\end{pmatrix}\begin{pmatrix} g_{\mathbf{1}}(z) \\ g_{\mathbf{Adj}}(z) \end{pmatrix}\equiv Og(z)
\end{align}
where $O$ is the orthogonal exchange matrix defined in subsection \ref{sec:crossing} and $g(\zb)$ transforms similarly. This means that the correlator transforms as
\begin{align}
g^T(z)Ag(\zb) ~\rightarrow~ g^T(z)O^TAOg(\zb).
\end{align}
Demanding invariance of the correlator amounts to the constraint $O^TAO=A.$ Any linear combination of the identity matrix and the exchange matrix $O$ satisfies this equation and will lead to a crossing symmetric correlation function when plugged in (\ref{pprcb}). We continue to compute these crossing symmetric building blocks. With $A=I$ we get
\begin{align}\label{gci}
\begin{split}
G_I(z_i,\zb_i)&=g_{\mathbf{1}}(z) g_{\mathbf{1}}(\zb) + g_{\mathbf{Adj}}(z) g_{\mathbf{Adj}}(\zb) \\
&=\frac{1}{N^2-1}\left[ \left(\vert z \vert^2\right)^{N-1}+\left(\vert z-1 \vert^2\right)^{N-1} - \frac{1}{N} ((z-1)\zb)^{N-1} - \frac{1}{N} (z(\zb-1))^{N-1}\right].
\end{split}
\end{align}
And with $A=O$ we get
\begin{align}\label{gco}
\begin{split}
G_O(z_i,\zb_i)&=\frac{1}{N}\big(g_{\mathbf{1}}(z)g_{\mathbf{1}}(\zb)-g_{\mathbf{Adj}}(z)g_{\mathbf{Adj}}(\zb)\big)+\frac{\sqrt{N^2-1}}{N}\big(g_{\mathbf{1}}(z)g_{\mathbf{Adj}}(\zb)+g_{\mathbf{Adj}}(z)g_{\mathbf{1}}(\zb)\big) \\
&=\frac{1}{N^2-1}\left[(z(\zb-1))^{N-1}+((z-1)\zb)^{N-1}-\frac{1}{N}\left(\vert z \vert^2\right)^{N-1}-\frac{1}{N}\left(\vert z-1 \vert^2\right)^{N-1}\right].
\end{split}
\end{align}
One can see in (\ref{pprxi})-(\ref{ppr}) that the exchange $x_2 \leftrightarrow x_4$ corresponds to $(z,\zb) \rightarrow (1-z,1-\zb)$ and (\ref{gci}) and (\ref{gco}) are indeed invariant under this transformation.
A specific linear combination of $G_I$ and $G_O$ gives
\begin{align}
G(z_i,\zb_i)=\left(\vert z \vert^2\right)^{N-1}+\left(\vert z-1 \vert^2\right)^{N-1}.
\end{align}
This is the semiclassical limit of the result computed in \cite{Papadodimas:2011pf} using the Coulomb gas formalism. Another linear combination of interest is the following
\begin{align}
G(z_i,\zb_i)= \left(\vert z \vert^2\right)^{N-1}+\left(\vert z-1 \vert^2\right)^{N-1} + ((z-1)\zb)^{N-1} + (z(\zb-1))^{N-1}.
\end{align}
For $N=-1$, this expression reduces to the following correlator of free complex bosons
\begin{align}
\begin{split}
G(z_i,\zb_i)&=\langle \partial \phi \bar{\partial} \overline{\phi} (x_1) \partial \overline{\phi} \bar{\partial} \phi (x_2) \partial \phi \bar{\partial} \overline{\phi} (x_3) \partial \overline{\phi} \bar{\partial} \phi (x_4) \rangle \\
&=(z_{12}\zb_{12})^{-2}(z_{34}\zb_{34})^{-2}+(z_{14}\zb_{14})^{-2}(z_{23}\zb_{23})^{-2} \\ 
&~~~~~~+(z_{12}\zb_{14})^{-2}(z_{34}\zb_{23})^{-2}+(z_{14}\zb_{12})^{-2}(z_{23}\zb_{34})^{-2}
\end{split}
\end{align}
after implementing coordinates as in (\ref{ccross}) and (\ref{ppr}).

\section{Heavy-light Virasoro blocks}

We now show how to use our approach to obtain Virasoro blocks in the heavy-light limit.  This refers to a limit in which we take  $c \rt \infty$ while scaling operator dimensions in a specific way.  In particular, we consider a four-point function of two light operators and two heavy operators, $\langle O_{L_1} O_{L_2} O_{H_1}O_{H_2}\rangle$. Light operators have scaling dimensions $h_{1,2}$ that are held fixed in the limit, while heavy operator dimensions $H_{1,2}$ scale like $c$, while their difference $H_{12}=H_1-H_2$ is held fixed.  Further, the exchanged primary is taken to be light, with its scaling dimension $h_p$ held fixed.

Rather than working on the z-plane, in this section it will be more convenient to work on the cylinder, $z=e^{iw}$, with $w=\phi+i\tau$.    Of course, the conformal blocks in the two cases are simply related by a conformal transformation.   We will further use conformal invariance to place the heavy operators in the far past and future, and one of the light operators at $w=0$.  With these comments in mind,
the heavy-light Virosoro blocks on the cylinder are \cite{Fitzpatrick:2015zha}
\bea
&&\langle   O_{L_1}(w,\wb) O_{L_2}(0,0)\, P_{p}\,  O_{H_1}(\tau=-\infty)O_{H_2}(\tau=\infty)\rangle =\F(h_i,h_p;w)\overline{\F}(\hb_i,\hb_p;\wb)
\cr &&
\eea
with
\bea
&&\F(h_i,h_p;w)=\left(\sin{\alpha w \over 2}\right)^{-2h_{L_1}} \left(1-e^{i\alpha w} \right)^{h_p+h_{12}} {_2{F_1}}\Big(h_p+h_{12},h_p-{H_{12}\over \alpha} ,2h_p;1-e^{i\alpha w} \Big)~.
\cr &&
\eea
Here
\be
\alpha =\sqrt{1 - {24 h_{H_1} \over c}}~.
\ee
Setting $\alpha=1$ yields the result for the global block.   We then note that the heavy-light Virasoro block is obtained from the global block by the replacements
\be\label{rescale}
w \rt \alpha w~,\quad H_{12} = {H_{12}\over \alpha}~.
\ee
We now show how this result comes out in our approach.

As shown in previous work, the relevant bulk geometry is a conical defect spacetime whose energy matches the dimension of the heavy operators.  The corresponding connection is
\be
 a = (L_1 +{\alpha^2 \over 4}L_{-1})dw
 \ee
We  now write
\be
 e^{(T_1 +{\alpha^2 \over 4}T_{-1})w}= e^{c_1(w)T_1}[c_0(w)]^{2T_0}e^{c_{-1}(w)T_{-1}}
 \ee
with
\be
 c_1(w) = {2\over \alpha}\tan {\alpha w\over 2}~,\quad c_0(w)= \cos {\alpha w\over 2}~,\quad c_{-1}(w)={\alpha\over 2}\tan{\alpha w\over 2}
 \ee
obtained by matching the two sides in the two-dimensional rep of SL(2).

The conformal block is given by\footnote{To avoid confusion with the cylinder coordinates $w_i$, we use $\mathfrak{w}_s$ to denote the conformal blocks in this section.}
\be
\mathfrak{w}_s(w_i) = \sum_{\{m_i\}}S_{m_1,m_2,m_3,m_4} \prod_{i=1}^4 \langle j_i m_i|  e^{c_1(w_{bi})T_1} [c_0(w_{bi})]^{2T_0}|j_ij_i\rangle
\ee
where we have written the singlet state as $\langle s|=\sum_{\{m_i\}}S_{m_1,m_2,m_3,m_4} \prod_{i=1}^4 \langle j_i m_i|$.
We will think of the first two spins as representing the light operators, so $h_1=-j_1$ and $h_2=-j_2$.  The insertion point of last two spins will be taken to $\tau= \pm \infty$, since this is where the heavy operators are inserted.  The heavy operators correspond to the background connection with the contribution of the spins added on top.  Below we will see that $H_{12}= -\alpha(j_3-j_4)$.

We use conformal invariance to set
 \be
 w_1 \rt w~,\quad w_2\rt 0~,\quad w_3 \rt -i\infty~,\quad w_4 \rt +i\infty~,\quad w_b \rt 0
 \ee
The functions behave as
\bea
&& c_1(w_{b2}) \sim 0~,\quad c_0(w_{b2}) \sim 1 \cr
&& c_1(w_{b3})\sim {2i\over \alpha}~,\quad c_0(w_{b3}) \sim  {1\over 2}e^{i\alpha w_3\over 2} \rt \infty\cr
&&c_1(w_{b4}) \sim - {2i\over \alpha}~,\quad c_0(w_{b4}) \sim  {1\over 2}e^{-{i\alpha w_4\over 2}} \rt \infty
\eea
The first limit  picks out $m_2=j_2$ from the sum.  After stripping off the $w_{3,4}$ dependent factors (which are absorbed into the definition of the operators at $\tau=\pm \infty$)  we are left with
\bea\label{gvres}
&& \mathfrak{w}_s(w_i) = (\cos {\alpha w\over 2})^{2j_1}\sum_{m_1,m_3,m_4} S_{m_1,j_2,m_3,m_4} \langle j_1 m_1 |  e^{-{2\over \alpha} \tan{\alpha w\over 2}T_1} |j_1j_1\rangle   \cr
 &&\quad\quad \quad\quad \quad\quad \quad\quad\quad\quad\quad\quad~ \langle j_3 m_3 |e^{{2i\over \alpha} T_1} |j_3j_3\rangle
  \langle j_4 m_4 | e^{-{2i\over \alpha}T_1}  |j_4j_4\rangle
\eea
Now, starting from the $\alpha=1$ case we obtain (\ref{gvres}) by the replacements
\be
 w\rt \alpha w~,\quad T_1 \rt {1\over \alpha}T_1
 \ee
We first establish that the rescaling of $T_1$ has no effect other than contributing an overall multiplicative constant.   This is because upon expanding the exponentials only a fixed overall power of $T_1$ contributes, since $m_1+j_2+m_3+m_4=0$ by the singlet condition.  We simply pick up one power of $\alpha$ for each power of $T_1$, which as noted above just yields a fixed overall constant which we ignore.

Besides the rescaling of $w$, we also need to account for the rescaling of $H_{12}$ in (\ref{rescale}).  At $\alpha=1$ we have only light operators and we would write $H_{12} = -(j_3-j_4)$.   For general $\alpha$ we can read off the contribution to the scaling dimension from $j_{3,4}$ from the $w_{3,4}$ dependent prefactor that we stripped off.  From the behavior of the functions  $c_0(w_{b3})$ and $c_0(w_{b4})$ we see that this factor is   $e^{i\alpha j_3 w_3 \over 2}e^{-{i\alpha j_3 w_4 \over 2}}$.  This tells us that it is $\alpha j_{3,4}$ that contributes to the scaling dimensions, and so $-(j_3-j_4)={H_{12}\over \alpha}$.  This accounts for the rescaling of $H_{12}$.

Altogether, we see that if we have established the correct result for the global conformal block, as we have indeed done in section \ref{SL2}, then agreement for the heavy-light block follows.  This completes the argument.

\section{Discussion}

We close with a few comments. The main result of this work is formula (\ref{master}), yielding large $c$ correlators and conformal blocks of ${\cal W}_N$ theories.  We showed by explicit computation how the choice of light external operators yields global blocks, recovering known results in a new way that is well adapted to holographic considerations.   We can equally well obtain heavy-light blocks, as was demonstrated in the $N=2$ case where we obtained heavy-light Virasoro blocks.   Similarly, heavy-light blocks for ${\cal W}_N$ can be obtained through more work, if desired.   In all these cases, all our results directly pertain to the case where operator dimensions are negative; however, after the result has been obtained one can analytically continue to positive dimensions.  Of course, this requires some knowledge of the analytic structure as a function of operator dimension.  This is usually no obstacle: for example, one knows that each term in the series expansion of a conformal block in the cross ratio is a rational function of operator dimensions, rendering analytic continuation trivial.   Similarly, one can analytically continue in $N$ to obtain blocks of ${\cal W}_{\infty}(\lambda)$.

Looking ahead, it would be very interesting to obtain (\ref{master}) directly from the equations of Prokushkin and Vasiliev.   At present, we only know how to do this in the case of two light operators, corresponding to computing a two-point function in a heavy background.
Starting from the Prokushkin-Vasiliev equations, it is well known (e.g. \cite{Ammon:2011ua}) how to linearize in the matter field to obtain a description of a free scalar interacting with Chern-Simons gauge fields, and how the computation of two-point functions leads to a special case of (\ref{master}).  However, the system of equations becomes much more complicated when matter self-interactions are included, and they have so far not been put into a usable form.
We also note that the case of two light operators includes all existing computations of entanglement entropy in higher spin theories, which correspond to two-point functions of operators with quantum numbers chosen to match those of twist operators \cite{deBoer:2013vca,Ammon:2013hba,Castro:2014mza,Hegde:2015dqh}.

Results obtained here pertain to the large $c$ limit, which corresponds to the classical limit in the bulk.   On the CFT side one can work out $1/c$ corrections \cite{Fitzpatrick:2015dlt}, and it is interesting to ask how these might arise in the bulk as quantum corrections.   For example, one might entertain computing loop diagrams in the bulk via Wilson lines.  However, the most obvious way of defining such diagrams does not lead to anything new when we recall that gauge invariance implies that the location of bulk vertices can be moved without changing the result.  The same argument that said that tree level exchange diagrams can be reduced to contact diagrams by merging vertices also tells us that such loop diagrams can be reduced to tree level contact diagrams.  Apparently some new ingredient is needed to compute quantum corrections.\newline

\vspace{.1in} 
\noindent
{ \bf \Large Acknowledgments}

\vspace{.1in}
 Work is supported in part by NSF grant PHY-1313986.


\appendix

\section{Conformal invariance of correlators}
\label{appconf}
Here we show that our correlation functions transform properly under global conformal transformations, as in (\ref{ae}).
We start from our general expression for an n-point function
\be\label{genex}
w_s(z_i) = \langle s| \prod_{i=1}^n  e^{z_{bi}T^{(i)}_1} |{\rm hw}\rangle_i~.
\ee
Under a gauge transformation of the connection
\begin{align}\label{a1a}
a \to LaL^{-1} + LdL^{-1}
\end{align}
a Wilson line  transforms as
\begin{align}\label{a1b}
P e^{\int_x^y a} \to L(y)Pe^{\int_x^y a}L^{-1}(x)
\end{align}
An arbitrary  $SL(2)$ transformation can be written as
\begin{align}\label{a1c}
L(z) = e^{\cm T_{-1}}~e^{2\log \cz T_0}~e^{\cp T_1}
\end{align}
where the $c_i$ are functions of $z$.  Starting with the connection corresponding to pure AdS in Poincar\'e coordinates , $a=T_1 dz$, a gauge transformation by $L(z)$ gives
\begin{align}\label{a1d}
a' = \left[\frac{1-\cp'}{\cz^2}T_1-\frac{2(\cm+\cz\cz'-\cm\cp')}{\cz^2}T_0-
\frac{\cz^2\cm'-2\cz\cz'\cm-\cm^2+\cm^2\cp'}{\cz^2}T_{-1}\right]dz
\end{align}
To verify this one can first work out the result in the $2\times 2$ matrix representation of $SL(2)$ and then use the fact that the group multiplication is independent of the representation. We demand that $a'\propto T_1$, so that the coefficients of $T_0$ and $T_{-1}$ vanish.  It will prove sufficient to take
\begin{align}\label{a1e}
c_1(z)=0~,\quad \cz(z) = cz+d~,~~\cm(z)=-c \cz(z)
\end{align}
corresponding to the new connection
\begin{align}\label{a1f}
a' = \frac{T_1 dz}{(cz+d)^2}=T_1 dz'
\end{align}
where
\be\label{a1g}
 z'={az+b\over cz+d}~,\quad ad-bc=1~.
\ee
Returning to (\ref{genex}) we write
\bea
 w_s(z_i)&=&\langle s| \prod_{i=1}^n L^{-1}(z_b)L(z_b) e^{z_{bi}T^{(i)}_1}L^{-1}(z_i)L(z_i) |{\rm hw}\rangle_i\cr
 & =&\langle s| \prod_{i=1}^n e^{z'_{bi}T^{(i)}_1}L(z_i) |{\rm hw}\rangle_i
 \eea
We further have
\begin{align}
L(z_i)\ket{{\rm hw}}_i &= e^{\cm T_{-1}}~e^{2\log \cz T_0}\ket{{\rm hw}}_i\cr
&= e^{-2h_i\log \cz}\ket{{\rm hw}}_i\cr
&= (cz_i+d)^{-2h_i}\ket{{\rm hw}}_i
\end{align}
which yields
\be
w_s(z_i) = \left[ \prod_{i=1}^n (cz_i+d)^{-2h_i} \right] w_s(z'_i)~.
\ee
This is equivalent to (\ref{ae}).

\section{Computation of three-point function}
\label{appthreepoint}
In this appendix we give the details for deriving (\ref{threepoint}).  We work with a description of SL(2) representations based on symmetric tensors, or equivalently Young tableau with a single row.
We start with Young tableaux with one single row of length $\lambda=2j$ for a spin $j$ representation. In tensor notation the states of this representation are $A^{{\alpha}_1 \dots {\alpha}_{\lambda}}\ket{e_{{\alpha}_1} \dots e_{{\alpha}_{\lambda}}}$ where $A$ is a symmetric tensor, and $\ket{e_1}$ and $\ket{e_2}$ are the spin up and spin down states of the spin half representation of SL(2), respectively. In other words, $\ket{e_1}$ and $\ket{e_2}$ are states in the fundamental representation of SL(2).  The highest weight state is $\ket{e_1 \dots e_1}$.
Wilson lines emanating from the boundary points $z_1, z_2$ and $z_3$ carry Dynkin labels $\lambda_1,\lambda_2$ and $\lambda_3$, respectively, and we take $\lambda_1 \geq \lambda_2$ without loss of generality. The tensor product of the first two representations decomposes as
\begin{align}
\Yvcentermath1
\begin{split}
\Yboxdim{20pt}\young(\lo)\otimes\young(\lt)&=\sum_{\lambda=\vert\lo-\lt\vert}^{\lo+\lt}\Yboxdim{20pt}\young(\lambda)
\end{split}
\end{align}
where  representations of label $\lambda\in \{|\lambda_1-\lambda_2|, \ldots, \lambda_1+\lambda_2\}$ appear. If $\lambda_3$ lies in this interval we can build a singlet out of the three representations. Once we have the singlet we need to evaluate the bulk-to-boundary Wilson lines. These act independently on each state of the fundamental representation so it is convenient to first evaluate matrix elements on these factors and then assemble the singlet, which will then lead directly to the three point function. We denote by $\qi_{\alpha}$ the following matrix element of the Wilson line
\begin{equation}\label{eq:q3ptsl2}
\qi_{\alpha}=\bra{e_{\alpha}}e^{z_{bi}T^{(i)}_1}\ket{e_1}_i=\delta^1_{\alpha}-z_{bi}\delta^2_{\alpha}.
\end{equation}
We now exploit gauge invariance to set $z_1=z_b$. After this we see that $\qo_{\alpha}=\delta^1_{\alpha}$, which simplifies the calculation.  We now define the tensor $(M_i)_{j_1\dots j_{\lambda_i}}=\langle ({\cal R}_i)_{j_1\dots j_{\lambda_i}}\vert e^{z_{bi}L_1} \vert \text{hw}\rangle_i$ representing the matrix element of the Wilson line for any state in the representation ${\cal R}_i$.
\begin{align}
\begin{split}
z_1 \quad &:  \quad (M_{1})_{\alpha_1\dots\alpha_{\lambda_1}}=\delta^{1}_{\alpha_1}\dots\delta^{1}_{\alpha_{\lambda_1}}\\
z_2 \quad &:  \quad (M_{2})_{\beta_1\dots\beta_{\lambda_2}}=\qt_{(\beta_1}\dots \qt_{\beta_{\lambda_2})}\\
z_3 \quad &:  \quad (M_{3})_{\rho_1\dots\rho_{\lambda_3}}=\qth_{(\rho_1}\dots \qth_{\rho_{\lambda_3})}\\
\end{split}
\end{align}
We now build a tensor of $\lambda_3$ symmetric indices out of $M_1$ and $M_2$.
\begin{align}
\begin{split}
(M_{12})_{{\gamma}_1\dots {\gamma}_{\lambda_3}}&=\epsilon^{\alpha_1\beta_1}\dots \epsilon^{\alpha_{{{\lambda_1+\lambda_2-\lambda_3}\over 2}} \beta_{{{\lambda_1+\lambda_2-\lambda_3}\over 2}}}(M_{1})_{\alpha_1\dots\alpha_{{{\lambda_1+\lambda_2-\lambda_3}\over 2}}(\gamma_1\dots \gamma_{{{\lambda_1+\lambda_3-\lambda_2}\over 2}}}(M_{2})_{\gamma_{{{\lambda_1+\lambda_3-\lambda_2}\over 2}+1}\dots \gamma_{\lambda_3})\beta_1\dots\beta_{{{\lambda_1+\lambda_2-\lambda_3}\over 2}}}\\
&=(\qt_2)^{{{\lambda_1+\lambda_2-\lambda_3}\over 2}} \delta^1_{(\gamma_1}\dots\delta^1_{\gamma_{{{\lambda_1+\lambda_3-\lambda_2}\over 2}}}\qt_{\gamma_{{{\lambda_1+\lambda_3-\lambda_2}\over 2}+1}}\dots \qt_{\gamma_{\lambda_3})}
\end{split}
\end{align}
where we have contracted indices with the invariant tensor $\epsilon^{\alpha \beta}$. Finally we bring $M_{12}$ and $M_3$ together and construct the singlet
\begin{align}
\begin{split}
w_s(z_1,z_2,z_3)&= \epsilon^{\alpha_1\beta_1}\dots \epsilon^{\alpha_{\lambda_3}\beta_{\lambda_3}}(M_{3})_{\alpha_1\dots \alpha_{\lambda_3}} (M_{12})_{\beta_1\dots \beta_{\lambda_3}} \\
&=(\qt_2)^{{{\lambda_1+\lambda_2-\lambda_3}\over 2}} \epsilon^{\alpha_1 \beta_1}\dots\epsilon^{\alpha_{\lambda_3} \beta_{\lambda_3}}\qth_{\alpha_1}\dots \qth_{\alpha_{\lambda_3}}  \delta^1_{\beta_1}\dots\delta^1_{\beta_{{{\lambda_1+\lambda_3-\lambda_2}\over 2}}}\qt_{\beta_{{{\lambda_1+\lambda_3-\lambda_2}\over 2}+1}}\dots \qt_{\beta_{\lambda_3}}
\end{split}
\end{align}
where we have made use of the symmetric structure of $M_3$ and $M_{12}$, and discarded constant factors. Using now the explicit form of $\qi_{\alpha}$ from (\ref{eq:q3ptsl2}) we obtain
\begin{equation}
w_s(z_1,z_2,z_3)=z^{{{\lambda_1+\lambda_2-\lambda_3}\over 2}}_{12}z^{{{\lambda_1+\lambda_3-\lambda_2}\over 2}}_{13}z^{{{\lambda_2+\lambda_3-\lambda_1}\over 2}}_{23}
\end{equation}
This yields the result (\ref{threepoint}) upon using $h_i=-\lambda_i/2$.

\section{SL(N) Conventions and Facts}\label{group theory}

\subsection{Conventions}\label{slnconventions}
We use the same conventions as in \cite{Castro:2011iw}. All the Wilson lines that appear in this paper are valued in the \SL{2} subgroup of \SL{N}. The matrices we then need are for the generators of \SL{2} which in the $N$ dimensional defining representation are
\al{gta}{
L_1 &= -
\begin{pmatrix}
0 &\ldots &&&&&0\\
\sqrt{N-1}&0&&&&\ldots&\\
0&\sqrt{2(N-2)}&0&&&&\\
\vdots&&\ddots&\ddots&&&\vdots\\
&&&\sqrt{i(N-i)}&0&&\\
&&&&\ddots&\ddots&\\
0&\ldots&&&&\sqrt{N-1}&0
\end{pmatrix}
\\
L_0 &= \text{diag}\left(\frac{N-1}{2}, \frac{N-3}{2},\ldots, \frac{N-2i+1}{2},\ldots,-\frac{N-1}{2}\right)
\\
L_{-1} &= -(L_1)^\dagger
}
Using these matrices, we find for the defining representation
\al{gtc}{
\braket{-\text{hw}|e^{z L_1}|\text{hw}} &= \frac{z^{N-1}}{(N-1)!}\braket{-\text{hw}|(L_1)^{N-1}|\text{hw}}
\\
&= (-z)^{N-1}
}

\subsection{Irreducible Tensors}
We denote states of the defining representation of \SL{N} by an n-dimensional lower indexed vector $\ket{e_i}$, $i=1,\ldots, N$. It is natural then to denote states of the conjugate representation by upper indexed objects $\ket{\bar{e}^i}$, such that the invariant tensors are given by $\delta_i^j$, $\epsilon_{i_1\ldots i_N}$ and $\epsilon^{j_1\ldots j_N}$. Their invariance follows from the fact that the matrices of \SL{N} have unit determinant.

This characterization is useful for the \SL{3} calculations of section \ref{sl3}. The invariant tensors are now $\delta_i^j$, $\epsilon_{ijk}$ and $\epsilon^{ijk}$. Consider a tensor with arbitrary number of lower and upper indices. First focus on a pair of lower indices. The part that is antisymmetric in these two indices can be converted into a single upper index using an $\epsilon^{ijk}$. Next we do the same with pairs of upper indices. We can keep doing this until we have a tensor that has completely symmetric upper and lower indices. We can also contract an upper and a lower index using $\delta_i^j$ to give a lower rank tensor. Thus an irreducible tensor of \SL{3} should be completely traceless and symmetric in upper and lower indices.

We can construct a symmetric traceless tensor with $m$ lower and $n$ upper indices, $T_{i_1\ldots i_m}^{j_1\ldots j_n}$, by taking a tensor product of $m$ copies of the defining and $n$ copies of its conjugate representation, symmetrizing and subtracting out traces. Since the traces are all lower rank tensors, we have
\al{gtb}{
\Yvcentermath1
\Yboxdim{20pt}\underbrace{\young(~)\otimes\cdots\otimes\young(~)}_m\otimes\underbrace{\young(~,~)\otimes\cdots\otimes\young(~,~)}_n \quad = \quad \young(nm,n)\oplus\cdots
}
where the $\ldots$ on the right denote Young tableau with boxes $< m+2n$. So we conclude $T_{i_1\ldots i_m}^{j_1\ldots j_n} \sim (m,n)$. Conjugation of a representation simply conjugates each factor in the tensor product above which is equivalent to exchanging upper and lower indices or $\overline{(m,n)} = (n,m)$.

\section{Removing Traces of Symmetric Tensors}\label{traceless}
Consider a tensor with $x$ lower indices and $y$ upper indices, $A_{i_1\cdots i_x}^{j_1\cdots j_y}$, where the upper and lower indices are completely symmetrized. For the purposes of having an irreducible representation of \SL{3} we also need this tensor to be traceless. Since all the indices are symmetric we can consider a particular trace, say $\delta_{j_1}^{i_1}$, all other traces being equivalent. The trace of the tensor $A_{i_1\cdots i_x}^{j_1\cdots j_y}$ gives a term with one upper and one lower index contracted --- a single trace expression. To make this tensor traceless, we need to subtract out single trace expressions with appropriate coefficients while maintaining the symmetry of the indices. Trace of the single trace terms we just added to our tensor gives new double trace expressions. We then subtract those double trace terms and keep on going until we run out of indices to contract. In general, the expression for the traceless tensor looks like
\al{a3a}{
\tilde{A}_{i_1\cdots i_x}^{j_1\cdots j_y} = A_{i_1\cdots i_x}^{j_1\cdots j_y} + \sum_{n=1}^{min(x,y)}C_n\delta_{(i_1}^{(j_1}\cdots\delta_{i_n}^{j_n}A_{i_{n+1}\cdots i_x)k_1\cdots k_n}^{j_{n+1}\cdots j_y)k_1\cdots k_n}
}
where the parentheses denote symmetrization. Our goal is then to fix the coefficients $\tilde{C}_n$. First note that since both $i$ and $j$ are symmetrized, a lot of the terms have the same tensor structure. For example, $\delta_{i_1}^{j_1}\delta_{i_2}^{j_2}$ is the same as $\delta_{i_2}^{j_2}\delta_{i_1}^{j_1}$ (but different from $\delta_{i_1}^{j_2}\delta_{i_2}^{j_1}$). To account for these degeneracies (given $n$), fix the indices that appear on the tensor $A$. There are $(x-n)!(y-n)!$ terms which are the same, coming from permutations of the $(x-n)$ lower and $(y-n)$ upper indices on $A$. Further, we have a total of $(n!)^2$ terms coming from the permutations of the lower and upper indices on the Kronecker deltas but only $n!$ of them are distinct corresponding to keeping the sequence of lower indices fixed while permuting the upper indices. This gives an additional degeneracy factor of $n!$. We then redefine our constants
\al{a3b}{
C_n = \frac{(-1)^n \tC_n}{(x-n)!(y-n)!n!}
}
such that each tensor structure appears with a factor of $(-1)^n\tC_n$ in the sum. Note that we have included a sign since the single trace terms cancel the zero trace terms, the double trace cancel the single trace terms and so on. We will use induction to determine $\tC_n$. Consider the term with $n$ traces and $n+1$ traces respectively.
\al{a3c}{
n &: \frac{(-1)^{n}\tC_{n}}{(x-n)!(y-n)!n!}\delta_{(i_1}^{(j_1}\cdots\delta_{i_n}^{j_n}A_{i_{n+1}\cdots i_x)k_1\cdots k_n}^{j_{n+1}\cdots j_y)k_1\cdots k_n}
\\
n+1 &: \frac{(-1)^{n+1}\tC_{n+1}}{(x-n-1)!(y-n-1)!(n+1)!}\delta_{(i_1}^{(j_1}\cdots\delta_{i_{n+1}}^{j_{n+1}}A_{i_{n+2}\cdots i_x)k_1\cdots k_{n+1}}^{j_{n+2}\cdots j_y)k_1\cdots k_{n+1}}
}
To facilitate counting, further restrict to a particular tensor structure after contracting with $\delta_{j_1}^{i_1}$, say $\delta_{i_2}^{j_2}\cdots\delta_{i_{n+1}}^{j_{n+1}}A_{i_{n+2}\cdots i_x k_1\cdots k_{n+1}}^{j_{n+2}\cdots j_y k_1\cdots k_{n+1}}$. This tensor structure can arise from the $n+1$ trace terms in one of 4 ways.
\begin{enumerate}
\item Both the indices $i_1$ and $j_1$ are among the Kronecker deltas and on the same Kronecker delta.
\al{a3d}{
\delta_{i_1}^{j_1}\delta_{i_2}^{j_2}\cdots\delta_{i_{n+1}}^{j_{n+1}}A_{i_{n+2}\cdots i_x k_1\cdots k_{n+1}}^{j_{n+2}\cdots j_y k_1\cdots k_{n+1}}
}
There is exactly one such term after accounting for the degeneracies. Contracting with $\delta_{i_1}^{j_1}$ gives an additional factor of 3.
\item Both the indices $i_1$ and $j_1$ are among the Kronecker deltas but are on different Kronecker deltas.
\al{a3e}{
\delta_{i_1}^{j_a}\delta_{i_2}^{j_2}\cdots\delta_{i_a}^{j_1}\cdots\delta_{i_{n+1}}^{j_{n+1}}A_{i_{n+2}\cdots i_x k_1\cdots k_{n+1}}^{j_{n+2}\cdots j_y k_1\cdots k_{n+1}}
}
where $2\leq a \leq n+1$. There are $n$ such terms and each gives a factor of 1.
\item $i_1$ is on a Kronecker delta but $j_1$ is on the tensor $A$.
\al{a3f}{
\delta_{i_1}^{j_b}\delta_{i_2}^{j_2}\cdots\delta_{i_{n+1}}^{j_{n+1}}A_{i_{n+2}\cdots i_x k_1\cdots k_{n+1}}^{j_{n+2}\cdots j_1\cdots j_y k_1\cdots k_{n+1}}
}
where $n+2\leq b\leq y$. There are $(y-n-1)$ such terms and each gives a factor of 1.
\item On a similar note, we can have $j_1$ on the Kronecker delta but $i_1$ on $A$.
\al{a3g}{
\delta_{i_c}^{j_1}\delta_{i_2}^{j_2}\cdots\delta_{i_{n+1}}^{j_{n+1}}A_{i_{n+2}\cdots i_1\cdots i_x k_1\cdots k_{n+1}}^{j_{n+2}\cdots j_y k_1\cdots k_{n+1}}
}
where $n+2\leq b\leq x$. There are $(x-n-1)$ such terms and each gives a factor of 1.
\end{enumerate}
It is easily checked that no other possibility gives the right tensor structure. Combining all this we get a factor of $(x+y-n+1)$ accompanying the required tensor structure in the $n+1$ trace terms. Looking at the terms in \eqref{a3c} with $n$ traces, the required tensor structure can appear only when both the $i_1$ and $j_1$ indices are on the tensor $A$ and the Kronecker deltas are in the correct form. This term occurs exactly once after removing degeneracies. Hence, we get the recursion relation
\al{a3h}{
\tC_{n+1} = \frac{\tC_n}{x+y-n+1}
}
Note that we can think of the original tensor as the $n=0$ term with $\tC_0 = 1$. The coefficients are then given by
\al{a3i}{
\tC_n=\frac{1}{[x+y+1]_n}
}
where $[a]_n$ is the descending Pochhammer symbol, $[a]_n = a(a-1)\ldots(a-n+1)$. Putting all of this together, the traceless tensor is given by
\al{a3j}{
\tilde{A}_{i_1\cdots i_x}^{j_1\cdots j_y} = \sum_{n=0}^{min(x,y)}\frac{(-1)^n\Gamma(x+y-n+2)}{\Gamma(x+y+2)\Gamma(x-n+1)\Gamma(y-n+1)\Gamma(n+1)}\delta_{(i_1}^{(j_1}\cdots\delta_{i_n}^{j_n}A_{i_{n+1}\cdots i_x)k_1\cdots k_n}^{j_{n+1}\cdots j_y)k_1\cdots k_n}
}

\section{Details of SL(3) Calculation}\label{combinatorics}

In this appendix, we present some details of the \SL{3} calculations of section \ref{sl3}. The singlet in terms of tensors of \SL{3} was found in \eqref{fn}. All that is required now is to contract all the indices while keeping track of all the combinatorial factors and powers of $z$.
\al{a4a}{
g_s(z) &= z^{2d}~\qo_{i_1}\cdots\qo_{i_{l_1}}\tq_{i_{l_1+1}}\cdots\tq_{i_x}\qbo^{j_1}\cdots\qbo^{j_{l_3}}\qbt^{j_{l_3+1}}\cdots\qbt^{j_{y}}
\\
&~\times\sum_{n=0}^{min(x,y)}C_nC'_n\delta^{(i_1}_{(j_1}\cdots\delta^{i_n}_{j_n}\qf_{j_{n+1}}\cdots\qf_{j_{n_4}}\tq'_{j_{n_4+1}}\cdots\tq'_{j_y}\qbf^{i_{n+1}}\cdots\qbf^{i_{n_1+n+1}}\qbth^{i_{n_1+n+2}}\cdots\qbth^{i_{x}}
}
As mentioned before we refer to the indices on the first line by colors and the second line by boxes. The various labels we use for indices and the number of them are collected below
\al{a4b}{
\begin{array}{lcccl}
\text{Label}&\quad\quad &\text{Index on}&\quad\quad &\multicolumn{1}{c}{\text{Number}}\\
\hline
\text{Red}&\quad\quad &\qo &\quad\quad &l_1=\lo-d\\
\text{Blue}&\quad\quad &\tq &\quad\quad &l_2=e\\
\text{Green}&\quad\quad &\qbo &\quad\quad &l_3=\lt-e\\
\text{Yellow}&\quad\quad &\qbt &\quad\quad &l_4=\mu-d-e\\
\text{Box 1}&\quad\quad &\qbf &\quad\quad &n_1=\ltp-e'\\
\text{Box 2}&\quad\quad &\qbth&\quad\quad &n_2=\mup-d'-e'\\
\text{Box 3/6}&\quad\quad &\delta &\quad\quad &n\\
\text{Box 4}&\quad\quad &\qf &\quad\quad &n_4=\lop-d'\\
\text{Box 5}&\quad\quad &\tq' &\quad\quad &n_5=e'\\
\end{array}
}
Each permutation will correspond to a particular way of coloring the boxes. Note that we are allowed to color boxes 1, 2 and 3 red or blue only and boxes 4, 5 and 6 green or yellow only. Taking this into account, the various contributions from different combinations are
\al{a4c}{
\begin{array}{lcccc}
\text{Coloring}&\quad\quad &\text{Contribution}&\quad\quad &\text{Number}\\
\hline
\text{Box 1 Red}&\quad\quad &\qo\cdot\qbf &\quad\quad &u\\
\text{Box 2 Red}&\quad\quad &\qo\cdot\qbth &\quad\quad &l_1-u-u'\\
\text{Box 3 Red}&\quad\quad &\qo_{j} &\quad\quad &u'\\
\text{Box 1 Blue}&\quad\quad &\tq\cdot\qbf &\quad\quad &n_1-u\\
\text{Box 2 Blue}&\quad\quad &\tq\cdot\qbth &\quad\quad &l_2-n_1-n+u+u'\\
\text{Box 3 Blue}&\quad\quad &\tq_{j} &\quad\quad &n-u'\\
\text{Box 4 Green}&\quad\quad &\qbo\cdot\qf &\quad\quad &v\\
\text{Box 5 Green}&\quad\quad &\qbo\cdot\tq' &\quad\quad &l_3-v-v'\\
\text{Box 6 Green}&\quad\quad &\qbo^{i} &\quad\quad &v'\\
\text{Box 4 Yellow}&\quad\quad &\qbt\cdot\qf &\quad\quad &n_4-n-v\\
\text{Box 5 Yellow}&\quad\quad &\qbt\cdot\tq' &\quad\quad &l_4-n_4+v+v'\\
\text{Box 6 Yellow}&\quad\quad &\qbt^{i} &\quad\quad &n-v'\\
\end{array}
}
Note that box 3 and box 6 must be contracted as they refer to lower and upper indices appearing on $\delta$. Our definition $\tq_j = \epsilon_{jbc}\qbo^b\qbt^c$ automatically gives $\tq\cdot \qbo = 0 = \tq\cdot \qbt$. Since $\Rc_1$ is represented as a symmetric traceless tensor, we also have $\qo\cdot \qbo = 0$. We are then left with just one possible combination -- color box 3 red and box 6 yellow giving a contribution of $\qo\cdot \qbt$. In the above table this means $u'=n$ and $v'=0$.

Choosing the bulk point to coincide with $z_1=0$ and imposing $z_4\to \infty$ constrains $\qo$, $\qbo$ to be highest weight ($\qo_1$ and $\qbo^3$) and $\qf$, $\qbf$ to be lowest weight ($\qf_3$ and $\qbf^1$). All the contributions can then be found by our knowledge of the matrix elements in the defining representation \eqref{fk}. For example we have
\al{a4d}{
\qo\cdot\qbth &= \qbth^1 = 1
\\
\tq\cdot\qbth &= \epsilon_{i3c}\qbth^i\qbt^c = \sqrt{2}z(1-z)
}
The next task is to find the combinatorial factors accompanying each combination and to sum them all up. As an example consider boxes of type 1 i.e. the first and fourth rows of table \eqref{a4c}. We need to color $u$ boxes red and the rest blue. First choose $u$ red indices and $n_1-u$ blue indices which can be done in ${l_1\choose u}{l_2\choose u}$ ways. The coloring of the $n_1$ boxes of type 1 can then be done in $\Gamma(n_1+1)$ ways. Proceeding in a similar manner with the rest of the boxes, we obtain
\al{a4e}{
g_p(z) &= z^{2d} \sum_{n=0}^{min(x,y)}C_nC'_n\sum_{u=0}^{n_1}\sum_{v=0}^{n_4-n}{l_1\choose u}{l_2\choose n_1-u}\Gamma(n_1+1){l_1-u\choose n}\Gamma(n+1)\Gamma(n_2-n+1)
\\
&\quad\quad\times{l_3\choose v}{l_4\choose n_4-n-v}\Gamma(n_4-n+1){l_4-n_4+n+v\choose n}\Gamma(n+1)\Gamma(n_5+1)
\\
&\quad\quad\times(\sqrt{2}z)^{n_1-u}(\sqrt{2}z(1-z))^{l_2-n_1+u}(-\sqrt{2})^{l_3-v}(-\sqrt{2}(1-z))^{l_4-n_4+v}z^{2n}
\\
&\sim z^{2d+l_2}\sum_{n=0}^{min(x,y)}C_nC'_n\frac{\Gamma(n_2-n+1)}{\Gamma(l_1-n+1)}z^{2n}(1-z)^{l_2-n_1+l_4-n_4}
\\
&\quad\quad\quad\quad\times {_2F}_1(-n_1,n-l_1;1+l_2-n_1;1-z){_2F}_1(-l_3,n-n_4;1+l_3-n_4;1-z)
}
where the $\sim$ indicates that we have ignored factors that are independent of $z$ and the summation variable $n$. We can put the hypergeometric functions into standard form using the identity
\al{a4f}{
{_2F}_1(a,b;b-m;z)=\frac{(-1)^m(a)_m}{(1-b)_m}(1-z)^{-a-m}{_2F}_1(-m,b-a-m;1-a-m;1-z)~,\quad m\in\mathbb{N}
}
where $(a)_m = a(a+1)\ldots(a+m-1)$ is the ascending Pochhammer symbol. We also use the following reflection formula for gamma functions
\al{a4g}{
\frac{\Gamma(s-a+1)}{\Gamma(s-b+1)} = (-1)^{b-a}\frac{\Gamma(b-s)}{\Gamma(a-s)}~,\quad a,b\in \mathbb{Z}, s\in \mathbb{C}
}
With $a=n$ and $b=0$, we obtain
\al{a4h}{
\Gamma(s-n+1) &= (-1)^{n}\frac{\Gamma(s+1)\Gamma(-s)}{\Gamma(-s+n)}
\\
&\sim \frac{(-1)^n}{(-s)_n}
}
The only other ingredient required is the factor $C_nC'_n$ which is obtained from \eqref{a3j} and \eqref{fm} to be
\al{a4i}{
C_nC'_n\sim \frac{(-1)^n}{\Gamma(n+1)}\frac{\Gamma(x+y-n+2)}{\Gamma(n_4-n+1)\Gamma(n_2-n+1)}
}
We then put all the factors and identities together into \eqref{a4e} and after the dust settles, we have
\al{a4j}{
g_p(z)\sim z^{2d+e}\sum_{n=0}^{min(x,y)}&\frac{z^{2n}}{n!}\frac{(-n_2)_n(-l_4)_n(-l_1)_n(-n_4)_n}{(-x)_n(-y)_n(-x-y-1)_n}
\\
&\times {_2F}_1(-l_2,n-n_2;n-x;z){_2F}_1(-n_5,n-l_4;n-y;z)
}
Note that we have the relations $n_1+n_2 = x = l_1+l_2$ and $n_4+n_5=y=l_3+l_4$ with all of the $l$'s and $n$'s being non-negative integers. We can then take the upper limit of the sum to be $\infty$ as all the extra terms in the sum vanish.

\bibliographystyle{ssg}
\bibliography{biblio}

\end{document}